\documentclass[aps,prd,floatfix,superscriptaddress,preprintnumbers]{revtex4}
\usepackage{amssymb}
\usepackage{amsmath}
\usepackage{amsfonts}
\usepackage{epsfig}             
\usepackage{graphicx}
\usepackage{tabularx}
\usepackage{bm}
\usepackage{color}
\newcommand{\seq}{\begin{subequations}}
\newcommand{\sen}{\end{subequations}}
\newcommand{\eq}{\begin{eqnarray}}
\newcommand{\en}{\end{eqnarray}}

\newcommand{\ra}{\rangle}
\newcommand{\bfn}{{\bf 0}_{\perp}}

\newcommand{\bfp}{{\bf p}_{\perp}}

\newcommand{\bfk}{{\bf k}_{\perp}}

\begin{document}

\title{New findings in gluon TMD physics} 

\author{Valery E. Lyubovitskij}
\affiliation{Institut f\"ur Theoretische Physik, Universit\"at T\"ubingen, 
Kepler Center for Astro \\ and Particle Physics, 
Auf der Morgenstelle 14, D-72076 T\"ubingen, Germany}
\affiliation{Departamento de F\'\i sica y Centro Cient\'\i fico
Tecnol\'ogico de Valpara\'\i so-CCTVal, Universidad T\'ecnica
Federico Santa Mar\'\i a, Casilla 110-V, Valpara\'\i so, Chile}
\author{Ivan Schmidt}
\affiliation{Departamento de F\'\i sica y Centro Cient\'\i fico
Tecnol\'ogico de Valpara\'\i so-CCTVal, Universidad T\'ecnica
Federico Santa Mar\'\i a, Casilla 110-V, Valpara\'\i so, Chile}

\date{\today}

\begin{abstract}

We revisit the model-independent decomposition of the gluon correlator, 
producing T-even and T-odd gluon transverse momentum distributions (TMDs), 
at leading twist. We propose an expansion of the gluon correlator, 
using a basis of four tensors (one antisymmetric and three symmetric), 
which are expressed through generators of the $U(2)$ group acting in the 
two-dimensional transverse plane. One can do clear interpretations of the 
two transversity T-odd TMDs with linear polarization of gluons: symmetric 
and asymmetric under permutation of the transverse spin of the nucleon and the
transverse momentum of the gluon. Using light-front wave function (LFWF) 
representation, we also derive T-even and T-odd gluon TMDs in the nucleon 
at leading twist. The gluon-three-quark Fock component in the nucleon 
is considered as bound state of gluon and three-quark core (spectator). 
The TMDs are constructed as factorized product of two LFWFs and gluonic matrix encoding 
information about both T-even and T-odd TMDs. In particular, T-odd TMDs arise due 
to gluon rescattering between the gluon and three-quark spectator. 
Gluon rescattering effects are parametrized by unknown scalar functions depending 
on the $x$ and $\bfk$ variables. 
Our gluon TDMs obey the model-independent Mulders-Rodrigues   
inequalities. We also derive new sum rules (SRs) involving T-even TMDs. 
One of the SRs states that the square of the unpolarized TMD 
is equal to a sum of the squares of three polarized TMDs. 
Based on the SR derived for T-even gluon TMDs, we make a conjecture 
that there should two additional SRs involving T-odd gluon TMDs, 
valid at orders $\alpha_s$ and $\alpha_s^2$. Then, we check these 
SRs at small and large values of $x$. We think that our study could serve 
as useful input for future phenomenological studies of TMDs.

\end{abstract}

\maketitle

\section{Introduction} 

Transverse momentum distributions (TMDs) of quarks and gluons 
play an essential role in the description of the three-dimensional picture
of hadrons. They encode their nonperturbative dynamics, 
depending on light-cone coordinate and transverse momentum of the  
parton, and can be extracted through analysis of QCD scattering processes.  
Gluon TMDs are proposed in Ref.~\cite{Mulders:2000sh} 
and were later considered in Refs.~\cite{Meissner:2007rx}-\cite{Lyubovitskij:2020xqj}.  
For reviews, see, e.g., Refs.~\cite{Boer:2011fh,Angeles-Martinez:2015sea}. 
In particular, in Ref.~\cite{Mulders:2000sh} the leading-twist gluon TMDs 
and fragmentation functions (FFs) have been studied in a model-independent way. 
For the first time in the literature, a decomposition of the leading-twist gluon tensors 
into gluon TMDs and FFs has been done. Moreover, model-independent positivity bounds involving 
T-even TMDs and FFs have been derived, using the condition of positivity of all 
the diagonal elements of the $2 \times 2$ submatrices, extracted from the $4 \times 4$ gluon 
$\otimes$ nucleon spin matrix. Then, in Ref.~\cite{Meissner:2007rx}, 
relations between gluon generalized parton distributions (GPDs) and TMDs were derived. 
At the same time, predictions for the GPDs and TMDs have been obtained, using 
two models --- a scalar diquark model~\cite{Brodsky:2002cx} and a quark target 
model~\cite{Meissner:2007rx}. In Ref.~\cite{Boer:2011kf}, the Boer-Mulders 
gluon TMD $h_1^{\perp g}(x,\bfk^2)$, characterizing the distribution of linearly polarized gluons 
in an unpolarized nucleon, was studied, assuming a simple Gaussian ansatz for 
the transverse momentum dependence $\bfk$ in $h_1^{\perp g}(x,\bfk^2)$ and the
unpolarized gluon TMD $f_1^g(x,\bfk^2)$ and taking into account the positivity bound 
$\bfk^2/(2M_N^2) \, |h_1^{\perp g}(x,\bfk^2)| \le f_1^g(x,\bfk^2)$, derived 
in Ref.~\cite{Mulders:2000sh}, where $x$ is the fraction 
of longitudinal momentum of hadron. In Ref.~\cite{Lorce:2013pza}, the
structure of the generalized off-diagonal and TMD quark-quark and gluon-gluon correlators 
for spin $\frac{1}{2}$ was analyzed, and  a parametrization in terms of the parton 
generalized TMDs for the quark-quark and gluon-gluon correlator was proposed. 
In the quark case it was consistent with a previously derived parametrization, done 
in terms of Lorentz structures, while in the case of gluons, it was presented 
for the first time in the literature. On the other hand, in Ref.~\cite{Boer:2015pni} 
three leading-twist T-odd gluon TMDs $x f_{1T}^{\perp g}(x,\bfk)$, 
$x h_{1T}^g(x,\bfk)$, and $x h_{1T}^{\perp g}(x,\bfk)$ 
inside a transversally polarized hadron at large $\bfk$ are 
calculated. There, it is shown that in the limit of small $x$, these three TMDs become equal 
and are determined by a spin-dependent Odderon~\cite{Zhou:2013gsa}.
Another study was done in Ref.~\cite{Boer:2016xqr}, where all leading-twist gluon TMDs were considered 
at small $x$. For the case of T-odd TMDs in a transversally polarized hadron, the results 
of Ref.~\cite{Boer:2015pni} were confirmed, while it was shown that the fourth TMD 
$g_{1T}^g(x,\bfk^2)$ in a transversally polarized hadron vanishes in that limit. 
It was also found that two TMDs $g_{1L}^g(x,\bfk^2)$ and $h_{1L}^{\perp g}(x,\bfk^2)$ inside 
a longitudinally polarized hadron also vanish at $x \to 0$, and in the case of two TMDs 
$f_1^g(x,\bfk^2)$ and $h_1^{\perp g}(x,\bfk^2)$, inside of the unpolarized nucleon, it was 
found that they are related at small $x$ as 
$\lim\limits_{x \to 0} \, x f_1^g(x,\bfk^2) = 
\bfk^2/(2M_N^2) \, \lim\limits_{x \to 0} \, x h_1^{\perp g}(x,\bfk^2)$. 
Furthermore, in Ref.~\cite{Lu:2016vqu}, the gluon Sivers TMD $f_{1T}^{\perp g}(x,\bfk^2)$ 
us calculated in a light-cone spectator model. 
A detailed analysis of T-even gluon TMD at leading twist has also been done in 
Ref.~\cite{Bacchetta:2020vty}, taking into account minimal and nonminimal 
couplings of the gluon with the three-quark spectators in nucleon. 
Moreover, in Ref.~\cite{Bacchetta:2018ivt}, the azimuthal asymmetries for $J/\psi$ and $\Upsilon$ 
production in deep-inelastic processes at leading order
in nonrelativistic QCD are studied,  in order to extract information about gluon TMDs. 
In Ref.~\cite{Kaur:2020pvc}, T-even gluon TMDs of the proton are calculated 
in a light-cone spectator model. 

A detailed analysis of gluon parton densities (PDFs, TMDs, and GPDs) 
and form factors is performed in Ref.~\cite{Lyubovitskij:2020xqj} 
using holographic QCD. There, it is found that the power behavior of gluon parton distributions
and form factors at large values of the light-cone variable and
large values of square momentum is
consistent with quark counting rules. It was also shown that the
transverse momentum distributions derived in holographic QCD 
obey the model-independent Mulders-Rodrigues (MR) inequalities, without
referring to specific model parameters.
All gluon parton distributions are defined in terms of the unpolarized
and polarized gluon parton distribution functions (PDFs). 
Based on analytical results for four T-even TMDs, 
a sum rule (SR) relating them was derived. From these SRs the 
MR inequalities~\cite{Mulders:2000sh} follow directly. 

The main objectives of the present paper are: (1) to reconsider a decomposition of 
the gluon leading-twist tensor on TMDs with special consideration of its transversal part; (2) to extend the
analysis done in Ref.~\cite{Lyubovitskij:2020xqj} to T-odd gluon TMD and to check the
consistency of the results at small $x$, with the model-independent results derived 
in Refs.~~\cite{Boer:2015pni,Boer:2016xqr}; (3) based on the SR derived 
in Ref.~\cite{Lyubovitskij:2020xqj} for T-even gluon TMDs valid at order $\alpha_s^0$, 
to make the conjecture that analogous SRs can be derived at orders $\alpha_s$ and 
$\alpha_s^2$, which involve also T-odd gluon TMDs. Then, we check these SRs 
at small and large values of $x$.  

The paper is organized as follows.
In Sec.~II, we present a different decomposition of gluon tensor 
at leading twist (especially its transverse part). 
In Sec.~III, we consider the derivation of gluon TMDs in 
light-front QCD motivated by soft-wall anti-de Sitter(AdS)/QCD. 
In Sec.~IV, we derive three SRs involving leading-twist gluon TMDs 
and check them with model-independent constraints at small $x$ 
and constraints dictated by constituent counting rules at large $x$.  
Finally, Sec.~V contains our conclusions.

\section{Gluon TMDs at leading twist}

\subsection{Kinematics and  notations}

In the following, we use light-cone kinematics, which is specified 
by two lightlike vectors $n_\pm$ as~\cite{Mulders:2000sh} 
\eq 
n_+^\mu = (1,0,\bfn)\,, \quad 
n_-^\mu = (0,1,\bfn) 
\en 
obeying the conditions $n_+ \, n_- = 1$ and $n_\pm^2 = 0$. 
Any 4-momentum $p$ is expanded through $n_\pm$ as 
\eq 
p^\mu = p^+ n_+^\mu + p^- n_-^\mu + p^\mu_\perp = (p^+,p^-,\bfp) \,. 
\en 
We work in the nucleon rest frame, where the nucleon $P$ 
and gluon $k$ momenta are specified as  
\eq 
P^\mu &=& \Big(P^+, P^-, \bfn \Big) \,, \quad P^- = \frac{M_N^2}{2P^+} \,, 
\nonumber\\
k^\mu &=& \Big(xP^+, \frac{k^2 + \bfk^2}{2xP^+}, \bfk\Big) \,, 
\en 
where $M_N$ is the nucleon mass and $x = k^+/P^+$ is the longitudinal 
momentum fraction carried by the gluon.  
The spin vector of the nucleon is expanded into one-dimensional 
longitudinal $S_L$ (helicity) and two-dimensional transverse ${\bf S}_T$ components 
in a manifestly covariant way, as
\eq 
S^\mu = S_L^\mu + S_T^\mu \,, 
\en 
where 
\eq 
S_L^\mu &=& S_L \, \frac{P n_-}{M_N} \, n_+^\mu 
        - S_L \, \frac{P n_+}{M_N} \, n_-^\mu 
        = S_L \, \biggl(\frac{P^+}{M_N}, - \frac{P^-}{M_N}, \bfn \biggr) \,, 
\nonumber\\
S_T^\mu &=& \Big(0, 0, {\bf S}_T\Big)      
\en 
with $S_L^2 + {\bf S}_T^2  = 1$. 
Note that in the infinite momentum frame, $P^+ \to \infty$, the $P^-$ component and the
components of all 4-vectors proportional to $P^-$ vanish and the 4-component spin vector 
is reduced to the three-component vector ${\bf S} = (S_L, {\bf S}_T)$ 
with $S_L = \cos\theta$ and ${\bf S}_T = (\cos\phi\sin\theta,\sin\phi\sin\theta)$, 
where $\theta$ and $\phi$ are the polar and azimuthal angles, 
defining an orientation of the spin-vector ${\bf S}$.   
In the following, we will also use the notation for rising/downward transverse spin 
$S_T^\pm = S_T^x \pm i S_T^y$. 

Next, we specify three symmetric, $g^{\mu\nu}_T$, 
$\eta^{\mu\nu}_T$, $\xi^{\mu\nu}_T$. and one 
antisymmetric, $\epsilon^{\mu\nu}_T$, transverse dimensionless 
tensors~\cite{Mulders:2000sh}: 
\eq 
g^{\mu\nu}_T &=& g^{\mu\nu} - n^\mu_+ n^\nu_- - n^\mu_- n^\nu_+ = 
\left( 
\begin{array}{cr}
0 & 0            \\[1mm]
0 & g^{ij}_T     \\ 
\end{array}
\right)    
\nonumber\\[3mm]
\eta^{\mu\nu}_T &=& g^{\mu\nu}_T - \frac{2 k^\mu_\perp k^\nu_\perp}{k_\perp^2} 
= \left( 
\begin{array}{cr}
0 &         0    \\[1mm]
0 & \eta^{ij}_T  \\ 
\end{array}
\right)    
\nonumber\\[3mm]
\xi^{\mu\nu}_T &=& \frac{e^{k_\perp \{ \mu}_T {k_\perp^\nu}^{ \} }}{k_\perp^2} 
= \left( 
\begin{array}{rc}
0 &                0\\[1mm]
0 & \xi^{ij}_T      \\ 
\end{array}
\right)    
\,, 
\nonumber\\[3mm]
\epsilon^{\mu\nu}_T &=& \epsilon^{\alpha\beta\mu\nu} \, n_{+ \alpha} \, n_{- \beta} 
= \epsilon^{n_+n_-\mu\nu} = \epsilon^{- + \mu\nu} 
= \left( 
\begin{array}{cr}
    0 & 0               \\
    0 & \epsilon^{ij}_T \\
\end{array}
\right)
\,, 
\en 
where we used the shortened notation involving the Levi-Civita 
tensor $\epsilon^{\mu\nu}$~\cite{Mulders:2000sh}:  
\eq
\epsilon^{A \{ \mu}_T {B^\nu}^{ \}} = 
\epsilon^{A \mu}_T B^\nu + 
\epsilon^{A \nu}_T B^\mu \,.
\en 
Contraction of the tensor $\epsilon^{\mu\nu}_T$ 
with four-component vectors $A_\mu$ and $B_\nu$ gives
\eq 
\epsilon^{A\nu}_T =
\epsilon^{\mu\nu}_T \, A_\mu\,, \quad 
\epsilon^{AB}_T =
\epsilon^{\mu\nu}_T \, A_\mu \, B_\nu \,.
\en 
In the case of the two-component vectors ${\bf S}_T$, $\bfk$, we have 
\eq 
\epsilon^{{\bf S}_T \bfk}_T =
\epsilon^{ij}_T \,  {\bf S}^i \, \bfk^j\,, \quad 
\epsilon^{\bfk i}_T =
\epsilon^{ji}_T \,  \bfk^j\,, \quad 
\epsilon^{{\bf S}_T i}_T =
\epsilon^{ji}_T \,  {\bf S}_T^j 
\,.
\en  
Here, $k^\mu_\perp = (0,0,\bfk) = \sqrt{\bfk^2} \, (0,0,\cos\phi_k,\sin\phi_k)$, 
with $k^2_\perp = - \bfk^2$, 
and $\phi_k$ is the azimuthal angle, defining an orientation of $\bfk$ in the 
transverse plane. $g^{ij}_T$, $\eta^{ij}_T$, $\xi^{ij}_T$, and 
$\epsilon^{ij}_T$ are the rank-2 matrices acting in the transverse plane, 
which are expressed through the generators of the $U(2)$ group 
(unit $I_{2\times 2}$ matrix and three Pauli matrices $\tau_i$, $i=1,2,3$) as 
\eq\label{U2_generators} 
g^{ij}_T &=& - \delta^{ij} = {\rm diag}(-1,-1)
\,,
\nonumber\\[3mm]
\eta^{ij}_T &=& \tau_3^{ij} \, \cos 2\phi_k 
+ \tau_1^{ij} \, \sin 2\phi_k = 
\left(
\begin{array}{cr}
    \cos 2\phi_k &   \sin 2\phi_k \\[1mm]
    \sin 2\phi_k & - \cos 2\phi_k \\
\end{array}
\right) 
\,, 
\nonumber\\[3mm]
\xi^{ij}_T &=& - \tau_3^{ij}  \, \sin 2\phi_k 
+ \tau_1^{ij}  \, \cos 2\phi_k = 
\left( 
\begin{array}{rc}
  - \sin 2\phi_k &   \cos 2\phi_k \\
    \cos 2\phi_k &   \sin 2\phi_k \\
\end{array}
\right)
\,, 
\nonumber\\[3mm]
\epsilon^{ij}_T &=& i \tau_2^{ij}  =  
\left( 
\begin{array}{cc}
  0 & 1 \\
 -1 & 0 \\
\end{array}
\right)
\,.
\en 
These four tensors ($g^{\mu\nu}_T$, $\eta^{\mu\nu}_T$, 
$\xi^{\mu\nu}_T$, and $\epsilon^{\mu\nu}_T$)   
and their mapping to the transverse plane ($g^{ij}_T$, $\eta^{ij}_T$, 
$\xi^{ij}_T$, and $\epsilon^{ij}_T$)   
play a fundamental role in the classification of TMDs~\cite{Mulders:2000sh} and obey 
the normalization and orthogonality conditions 
\eq 
L^{\mu\nu}_T  \  L_{\mu\nu, T} = 2\,, \qquad 
L^{\mu\nu}_T  \  L^\prime_{\mu\nu, T} = 0   
\en 
and 
\eq 
L^{ij}_T  \  L^{ij}_T  =  2 \,, \qquad 
L^{ij}_T  \  L^{\prime\ ij}_T  =  0 \,. 
\en 
for $L, L^\prime = g, \eta, \xi, \epsilon$ and $L \neq L^\prime$.

Another interesting property following from the well-known 
identity 
$\tau_i \tau_j = \delta_{ij} + i \epsilon_{ijk} \tau_k$ is  
\eq 
& &    (\eta_T \, \epsilon_T)^{ij} 
= - (\epsilon_T \, \eta_T)^{ij} 
= \xi_T^{ij}\,, \nonumber\\
& &    (\epsilon_T \, \xi_T)^{ij} 
= - (\xi_T \, \epsilon_T)^{ij} 
= \eta_T^{ij}\,, 
\\
& &    (\eta_T \, \xi_T)^{ij} 
=    - (\xi_T \, \eta_T)^{ij} 
= \epsilon_T^{ij}\,.  
\nonumber
\en   

Note that in the paper~\cite{Mulders:2000sh} five tensors were used
for the classification of gluon TMD: the four tensors 
discussed above plus the fifth symmetric tensor $\omega_T^{\mu\nu}$, 
which is a linear combination of two symmetric tensors 
$\eta_T^{\mu\nu}$ and $\xi_T^{\mu\nu}$: 
\eq\label{wT_subs} 
\omega^{\mu\nu}_T \equiv 
- \Big[ \epsilon^{k_\perp \{ \mu} \, {S^\nu_T}^{\}} + 
        \epsilon^{S_T     \{ \mu} \, {k^\nu_\perp}^{\}} \Big] = 
2 \, \Big[ \xi_T^{\mu\nu} \, {\bf S}_T \bfk 
+ \eta_T^{\mu\nu} \, e^{{\bf S}_T \bfk} \Big] 
= \left( 
\begin{array}{cr}
    0 & 0               \\
    0 & \omega^{ij}_T \\
\end{array}
\right)
\,, 
\en 
where 
\eq 
\omega^{ij}_T = 2 \, |{\bf S}_T| \, |\bfk| \, 
\left( 
\begin{array}{rc}
  - \sin\delta & \cos\delta \\
    \cos\delta & \sin\delta \\
\end{array}
\right)
\,, \qquad \delta = \phi+\phi_k \,.  
\en 
One can see that the exclusion of the fifth 
tensor $\omega^{ij}_T$ from the expansion of gluon correlator into TMDs 
has several advantages: 

(i) We reduce the number of tensors 
involved in the expansion of the gluon correlator on TMDs. 

(ii) The tensor $\omega^{\mu\nu}_T$ involves transverse spin, while the other four 
tensors ($g^{\mu\nu}_T$, $\eta^{\mu\nu}_T$, $\xi^{\mu\nu}_T$, 
and $\epsilon^{\mu\nu}_T$) are manifestly independent on $S_T$. 

(iii) The substitution of $\omega^{\mu\nu}_T$ into the gluon correlator, as a linear combination 
of $\eta^{\mu\nu}_T$ and $\xi^{\mu\nu}_T$, gives a clear separation of the T-odd transversity 
TMD, describing the distribution of linearly polarized gluons in the transversally  
polarized nucleons, into two terms. One of them standing in front of 
$\xi^{\mu\nu}_T \, {\bf S}_T \bfk$ structure, which is manifestly symmetric under 
${\bf S}_T \leftrightarrow \bfk$ interchange, 
we name symmetric transversity TMD $h_{1T}^{+ g}(x,\bfk^2)$, 
while the second TMD standing in front of antisymmetric structure 
$\eta^{\mu\nu}_T \, e^{{\bf S}_T \bfk}$ 
with respect to ${\bf S}_T \leftrightarrow \bfk$  we name 
antisymmetric transversity TMD $h_{1T}^{- g}(x,\bfk^2)$. 

The gluon polarization vectors $\epsilon^\mu_{\lambda}$ read
\eq
\epsilon^\mu_{\pm} = (0,0,{\bm\epsilon}_\pm) =
\frac{1}{\sqrt{2}} \, (0,0,\mp 1,-i)
\quad
\en
in the case of circular polarization and
\eq
\epsilon^\mu_{x}
= (0,0,{\bm\epsilon}_x) = (0,0,1,0)\,, \quad
\epsilon^\mu_{y} = (0,0,{\bm\epsilon}_y) = (0,0,0,1)\,.
\en
in the case of linear polarization.
The two sets of polarization vectors are related as:
\eq
\epsilon^\mu_\pm = \mp \frac{1}{\sqrt{2}} \,
\biggl[ \epsilon^\mu_x \pm i \epsilon^\mu_y
\biggr]
\,.
\en

We also define a covariant expression for the gluon 
spin density matrix~\cite{Berestetskii:2012be}: 
\eq 
\rho_{\lambda\lambda'}^{\mu\nu} = \epsilon_{\lambda}^\mu \, \epsilon_{\lambda'}^{\dagger \nu} \,. 
\en 
For both choices of polarization vectors, $\epsilon_{\lambda}^\mu$ 
obey the completeness and orthonormality
conditions:
\eq\label{gmunu_matrix}
- g^{\mu\nu}_T = \sum\limits_{\lambda\lambda'} \, 
\delta_{\lambda\lambda'} \, \rho_{\lambda\lambda'}^{\mu\nu}
= \sum\limits_{\lambda}
\, \epsilon^\mu_\lambda  \, \epsilon^{\dagger\nu}_\lambda \,, 
\qquad\qquad 
g_{\mu\nu; T} \, \rho_{\lambda\lambda'}^{\mu\nu} = 
\epsilon^{\dagger\mu}_{\lambda'} \, \epsilon_{\mu\lambda} 
= - \delta_{\lambda\lambda'} \,.
\en
The antisymmetric tensor $\epsilon^{\mu\nu}_T$ is expressed in terms of
the polarization vectors, 
for the case of linear and circular polarizations, as
\eq\label{emunu} 
- i \epsilon^{\mu\nu}_T = - i \, \Big[
\epsilon^{\mu}_x \, \epsilon^{\dagger\nu}_y -
\epsilon^{\mu}_y \, \epsilon^{\dagger\nu}_x \Big]
= \epsilon^{\mu}_+ \, \epsilon^{\dagger\nu}_+
- \epsilon^{\mu}_- \, \epsilon^{\dagger\nu}_- \,.
\en
In matrix form, Eq.~(\ref{emunu}) reads
\eq\label{emunu_matrix} 
- i \epsilon^{\mu\nu}_T 
= -i \sum\limits_{\lambda,\lambda'=x,y} 
\, \rho^{\mu\nu}_{\lambda\lambda'} 
\, \epsilon_T^{\lambda\lambda'} 
= \sum\limits_{\lambda,\lambda'=+,-} 
\, \rho^{\mu\nu}_{\lambda\lambda'} 
\, \tau_3^{\lambda\lambda'} 
\,. 
\en 

The symmetric tensors $\eta^{\mu\nu}_T$ and $\xi^{\mu\nu}_T$ are expressed in terms of 
gluon polarization vectors as 
\eq 
\eta^{\mu\nu}_T &=& 
\sum\limits_{\lambda,\lambda'=x,y} 
\, \rho^{\mu\nu}_{\lambda\lambda'} 
\, \eta_T^{\lambda\lambda'} 
= \sin 2\phi_k \, \Big[ 
\epsilon^{\mu}_x \epsilon^{\dagger\nu}_y
+ 
\epsilon^{\mu}_y \epsilon^{\dagger\nu}_x 
\Big] 
\,+\,  
\cos 2\phi_k \, \Big[ 
\epsilon^{\mu}_x \epsilon^{\dagger\nu}_x
- 
\epsilon^{\mu}_y \epsilon^{\dagger\nu}_y
\Big] 
\,, \\
\xi^{\mu\nu}_T &=& 
\sum\limits_{\lambda,\lambda'=x,y} 
\, \rho^{\mu\nu}_{\lambda\lambda'} 
\, \xi_T^{\lambda\lambda'} 
= \cos 2\phi_k \, \Big[ 
\epsilon^{\mu}_x \epsilon^{\dagger\nu}_y
+
\epsilon^{\mu}_y \epsilon^{\dagger\nu}_x
\Big]
\,-\, 
\sin 2\phi_k \, \Big[ 
\epsilon^{\mu}_x \epsilon^{\dagger\nu}_x
- 
\epsilon^{\mu}_y \epsilon^{\dagger\nu}_y
\Big] 
\en 
for the case of linear gluon polarization and 
\eq
\eta^{\mu\nu}_T &=& 
\sum\limits_{\lambda,\lambda'=+,-} 
\, \rho^{\mu\nu}_{\lambda\lambda'} 
\, \eta^{\lambda\lambda'} 
= - \Big[ 
  e^{-2i\phi_k} \, 
\epsilon^{\mu}_+ \epsilon^{\dagger\nu}_- 
+ e^{2i\phi_k} \, 
\epsilon^{\mu}_- \epsilon^{\dagger\nu}_+ 
\Big] 
\,, \label{eta_circ}\\
\xi^{\mu\nu}_T &=& 
\sum\limits_{\lambda,\lambda'=+,-} 
\, \rho^{\mu\nu}_{\lambda\lambda'} 
\, \xi^{\lambda\lambda'} 
= 
i \, \Big[ e^{-2i\phi_k} \, 
\epsilon^{\mu}_+ \epsilon^{\dagger\nu}_-
- 
e^{2i\phi_k} \, 
\epsilon^{\mu}_- \epsilon^{\dagger\nu}_+ \Big] 
\label{xi_circ}
\en 
for the case of circular gluon polarization. 

The matrices $\eta^{\lambda\lambda'}$ and $\xi^{\lambda\lambda'}$ 
are linear combinations of Pauli matrices $\tau_1$ and $\tau_2$: 
\eq 
\eta = 
- \cos 2\phi_k \, \tau_1 \,-\, 
  \sin 2\phi_k \, \tau_2 \,, \qquad 
\xi = 
\sin 2\phi_k \, \tau_1 \,-\, 
\cos 2\phi_k \, \tau_2 \,. 
\en 
Note that the sets of the $2 \times 2$ matrices, which define the expansion of 
the tensors $\epsilon^{\mu\nu}_T$, and $\eta^{\mu\nu}_T$, $\xi^{\mu\nu}_T$, 
are different in the cases of the linear 
($\epsilon^{\lambda\lambda'}_T$, 
$\eta^{\lambda\lambda'}_T$, and $\xi^{\lambda\lambda'}_T$) and circular 
($\tau^{\lambda\lambda'}_3$, $\eta^{\lambda\lambda'}$, and $\xi^{\lambda\lambda'}$) 
polarizations. 

\subsection{Expansion of the gluon correlator into TMDs}

We start from the well-known result derived by Mulders and Rodrigues~\cite{Mulders:2000sh} 
for the  gluon correlator expansion at leading twist, into the set of TMDs. Following the discussion 
in the previous section, we use the set of the four linear-independent matrices $g^{ij}_T$, $\eta^{ij}_T$, 
$\xi^{ij}_T$, and $\epsilon^{ij}_T$, acting in the two-dimensional (2D) transverse space, 
which are expressed through generators of the group $U(2)$ via relations~(\ref{U2_generators}). 
Note that we can proceed in four-dimensional Lorentz space, but for simplicity, we map the gluon tensor 
into the 2D transverse plane. 
 
The gluon correlators $\Gamma^{ij}(x,\bfk,{\bf S})$ split into the three terms 
[unpolarized $\Gamma^{ij}_U(x,\bfk,{\bf S})$,  
longitudinally polarized $\Gamma^{ij}_L(x,\bfk,{\bf S})$, and  
transversally polarized  $\Gamma^{ij}_T(x,\bfk,{\bf S})$] as~\cite{Mulders:2000sh} 
\eq 
\Gamma^{ij}(x,\bfk, {\bf S}) = \sum\limits_{P=U,L,T} \, \Gamma^{ij}_P(x,\bfk,{\bf S}) \,. 
\en 
In our analysis, for simplicity,
we drop the overall factor $x/2$ in the definition of $\Gamma^{ij}(x,\bfk, {\bf S})$. 
The unpolarized term $\Gamma^{ij}_U$ is expanded in terms of two TMDs --- 
$f_1^g(x,\bfk^2)$ (unpolarized gluon TMD) and 
$h_1^{\perp g}(x,\bfk^2)$ (linearly polarized gluon TMD) 
in unpolarized nucleon ---  as 
\eq
\Gamma^{ij}_U(x,\bfk,{\bf S}) = 
- g^{ij}_T \, f_1^g(x,\bfk^2) + \eta^{ij}_T \, h_1^{(1) \perp g}(x,\bfk^2) \,. 
\en 
Here, for the $h_1^{\perp g}$ TMD and in the following for the others, 
we use the notation, following Ref.~\cite{Mulders:2000sh}, 
\eq\label{TMD_even} 
{\rm TMD}^{(n)}(x,\bfk^2) = \biggl[\frac{\bfk^2}{2M_N^2}\biggr]^n \, {\rm TMD}(x,\bfk^2) \,.
\en  
The longitudinally polarized term $\Gamma^{ij}_L$ is expanded in terms of two TMDs ---
$g_{1L}^g(x,\bfk^2)$ (circularly polarized gluon TMD) and 
$h_{1L}^{\perp g}(x,\bfk^2)$ (linearly polarized gluon TMD) 
in the longitudinally polarized nucleon:  
\eq 
\Gamma^{ij}_L(x,\bfk,{\bf S}) = 
- i \epsilon^{ij}_T \, S_L \, g_{1L}^g(x,\bfk^2) 
+ \xi^{ij}_T \, S_L \, h_{1L}^{(1) \perp g}(x,\bfk^2) \,. 
\en 
The transversally polarized term $\Gamma^{ij}_T$ is expanded in terms of four TMDs ---
$f_{1T}^{\perp g}(x,\bfk^2)$ (unpolarized gluon TMD), 
$g_{1T}^g(x,\bfk^2)$ (circularly polarized gluon TMD), and 
$h_{1T}^{+ g}(x,\bfk^2)$ and $h_{1T}^{- g}(x,\bfk^2)$ 
(linearly polarized gluon TMDs) in transversally polarized nucleon --- as 
\eq 
\Gamma^{ij}_T(x,\bfk,{\bf S}) &=&  
          - g^{ij}_T \, \frac{e^{{\bf S}_T \bfk}}{M_N} \, f_{1T}^{\perp g}(x,\bfk^2) 
 - i \epsilon^{ij}_T \, \frac{{\bf S}_T \bfk}{M_N}     \, g_{1T}^g(x,\bfk^2) 
\nonumber\\
      &+&  \xi^{ij}_T \, \frac{{\bf S}_T \bfk}{M_N}     \, h_{1T}^{+ g}(x,\bfk^2) 
       + \eta^{ij}_T \, \frac{e^{{\bf S}_T \bfk}}{M_N} \, h_{1T}^{- g}(x,\bfk^2) 
\,.
\en

It is necessary to make a few comments regarding the correspondence 
of our set of gluon TMDs to the ones introduced in 
Ref.~\cite{Mulders:2000sh} and used 
in Refs.~\cite{Boer:2015pni,Boer:2016xqr}. 
First of all, we use the relation~(\ref{wT_subs}) to express tensor $\omega^{ij}_T$ 
via linear combination of tensors $\xi^{ij}_T$ and $\eta^{ij}_T$ and  
introduce two linearly polarized gluon TMDs $h_{1T}^{\pm g}(x,\bfk^2)$, 
which are related to the corresponding Mulders-Rodrigues TMDs --- 
linearity $\Delta H_T(x,\bfk^2)$ 
and pretzelosity $\Delta H_T^\perp(x,\bfk^2)$~\cite{Mulders:2000sh} --- as 
\eq 
h_{1T}^{\pm g}(x,\bfk^2) = - \frac{1}{2} \biggl[ 
\Delta H_T(x,\bfk^2) \pm \frac{\bfk^2}{2M_N^2} \Delta H_T^\perp(x,\bfk^2) 
\biggr] \,.
\en 
One can see that the  $h_{1T}^{+ g}(x,\bfk^2)$ and $h_{1T}^{- g}(x,\bfk^2)$ TMDs 
accompany the symmetric ${\bf S}_T \bfk$ and antisymmetric $e^{{\bf S}_T \bfk}$
correlation structures, 
under interchange of ${\bf S}_T$ and $\bfk$. 
Our transverse TMDs $h_{1T}^{\pm g}$ are related with 
the analogous sets $(h_{1T}^g$, $h_{1T}^{\perp g})$ and 
$(h_1$, $h_{1T}$, $h_{1T}^{\perp})$ used in Refs.~\cite{Boer:2015pni} 
and~\cite{Boer:2016xqr}, respectively, as: 
\eq 
h_{1T}^{+ g}(x,\bfk^2) &=& h_{1T}^g(x,\bfk^2) - h_{1T}^{\perp g}(x,\bfk^2) 
= \frac{1}{2} \biggl[ h_1(x,\bfk^2) + \frac{\bfk^2}{2M_N^2} h_{1T}^{\perp}(x,\bfk^2) \biggr]\,, 
\nonumber\\
h_{1T}^{- g}(x,\bfk^2) &=& h_{1T}^g(x,\bfk^2) 
= \frac{1}{2} \biggl[ h_1(x,\bfk^2) - \frac{\bfk^2}{2M_N^2} h_{1T}^{\perp}(x,\bfk^2) \biggr] 
= \frac{1}{2} h_{1T}(x,\bfk^2) \,. 
\en  
Our Sivers function $f_{1T}^{\perp g}(x,\bfk^2)$ differs by a sign from the 
corresponding function $G_T(x,\bfk^2)$ introduced in Ref.~\cite{Mulders:2000sh} 
and has the same sign as in Refs.~\cite{Meissner:2007rx,Boer:2015pni,Boer:2016xqr}.  

Now we present the decomposition of the full gluon correlation tensor into the more compact form: 
\eq\label{Gluon_Tensor_Full} 
\Gamma^{ij}(x,\bfk,{\bf S}) = 
- g^{ij}_T \, F_1^g(x,\bfk^2;{\bf S}_T) 
- i \epsilon^{ij}_T \, {\bf S} \, {\bf G}_1^g(x,\bfk^2) 
+ \eta^{ij}_T \, H_1^{(\eta) g}(x,\bfk^2;{\bf S}_T) 
 +   \xi^{ij}_T \, {\bf S} \, {\bf H}_1^{(\xi) g}(x,\bfk^2) \,, 
\en  
where 
\eq
F_1^g(x,\bfk^2;{\bf S}_T) &=& f_1^g(x,\bfk^2) 
+ \frac{e^{{\bf S}_T \bfk}}{M_N} \, f_{1T}^{\perp g}(x,\bfk^2) \,, 
\nonumber\\
{\bf G}_1^g(x,\bfk^2) &=& \biggl(g_{1L}^g(x,\bfk^2),\frac{\bfk}{M_N} g_{1T}^g(x,\bfk^2)\biggr) \,,
\nonumber\\
H_1^{(\eta) g}(x,\bfk^2;{\bf S}_T) &=& h_1^{(1)\perp g}(x,\bfk^2) 
+ \frac{e^{{\bf S}_T \bfk}}{M_N} \, h_{1T}^{- g}(x,\bfk^2) \,, 
\nonumber\\
{\bf H}_1^{(\xi) g}(x,\bfk^2) &=& \biggl(h_{1L}^{(1)\perp g}(x,\bfk^2),  
\frac{\bfk}{M_N} \, h_{1T}^{+ g}(x,\bfk^2)\biggr) \,. 
\en 
The terms in the gluon tensor expansion~(\ref{Gluon_Tensor_Full}) are fixed 
by contraction with the corresponding tensor acting in 2D transverse space, as: 
\eq 
F_1^g(x,\bfk^2;{\bf S}_T) &=& - \frac{g^{ij}_T}{2} 
\ \Gamma^{ij}(x,\bfk,{\bf S}) \,, \nonumber\\
{\bf S} \, {\bf G}_1^g(x,\bfk^2) &=& 
\frac{i\epsilon^{ij}_T}{2} \ \Gamma^{ij}(x,\bfk,{\bf S}) \,, \nonumber\\
H_1^{(\eta) g}(x,\bfk^2;{\bf S}_T) &=& \frac{\eta^{ij}_T}{2} 
\ \Gamma^{ij}(x,\bfk,{\bf S}) \,, \nonumber\\
{\bf S} \, {\bf H}_1^{(\xi) g}(x,\bfk^2) &=& 
\frac{\xi^{ij}_T}{2} 
\ \Gamma^{ij}(x,\bfk,{\bf S}) \,. 
\en  
 
Notice that the components of the gluon tensor $\Gamma^{ij}(x,\bfk)$, in the 
circular $i,j = +, -$ and linear $i,j = 1, 2$ basis, are related as~\cite{Mulders:2000sh} 
\eq 
\Gamma^{\pm\pm} &=&   \frac{1}{2} \, \Big(\Gamma^{11} + \Gamma^{22}\Big) 
\, \mp \, {\rm Im}(\Gamma^{12}) \,, \nonumber\\
\Gamma^{\pm\mp} &=& - \frac{1}{2} \, \Big(\Gamma^{11} - \Gamma^{22}\Big) 
\, \pm \, i {\rm Re}(\Gamma^{12}) \,. 
\en 
Following Ref.~\cite{Mulders:2000sh}, we get explicit results for the 
matrix elements $\Gamma^{ij}(x,\bfk)$ in the circular basis: 
\eq 
\Gamma^{\pm\pm} &=& f_1^g \pm S_L \, g_{1L}^g 
\pm \frac{|\bfk|}{2M_N} \, \Big(
e^{i\phi_k}  \, S_T^- \, 
\Big[g_{1T}^g \mp i f_{1T}^{\perp g}\Big] \, + \, 
e^{-i\phi_k} \, S_T^+ \, 
\Big[g_{1T}^g \pm i f_{1T}^{\perp g}\Big] 
\Big) 
\,, \nonumber\\
\Gamma^{\pm\mp} &=& e^{\mp 2i\phi_k} \, 
\Big(-h_1^{(1) \perp g}  \pm i S_L \, h_{1L}^{(1) \perp g}\Big) 
\pm \frac{i |\bfk|}{2 M_N} \, \Big( 
e^{\mp i\phi_k}   \, S_T^\mp \, \Big[h_{1T}^{+ g} +  h_{1T}^{- g}\Big] 
\, + \,  
e^{\mp 3 i\phi_k} \, S_T^\pm \, \Big[h_{1T}^{+ g} -  h_{1T}^{- g}\Big] 
\Big) \,. 
\en 
Next, we expand the gluon tensor in the nucleon spin basis~\cite{Mulders:2000sh} 
\eq 
\Gamma^{ij}(x,\bfk,{\bf S}) = \sum\limits_{\Lambda,\Lambda'} 
\, \rho_{\Lambda'\Lambda}({\bf S}) \, \Gamma^{ij}_{\Lambda\Lambda'}(x,\bfk) \,, 
\en 
where 
\eq 
\rho({\bf S}) = \frac{1}{2} \, \Big({\bf 1} + {\bf S} \, {\bf\sigma}\Big)
\en 
is the spin-$\frac{1}{2}$ density matrix and $\sigma^i$ is the triplet of the Pauli 
matrices corresponding to the nucleon spin. 
For convenience in our paper, we use two notations for Pauli matrices -- $\tau^i$, which are relevant
for the spin structure of gluon and $\sigma^i$ related to the spin structure of nucleon.

The density 
$\Gamma^{ij}_{\Lambda\Lambda'}(x,\bfk) = 
\rho_{\Lambda'\Lambda}({\bf S}) \, \Gamma^{ij}(x,\bfk,{\bf S})$
is the $4 \otimes 4$ matrix 
in the gluon $\otimes$ nucleon spin space, which is given by  
\eq 
\Gamma^{ij}_{\Lambda\Lambda'}(x,\bfk) &=& 
- g^{ij}_T \, 
\Big(\delta_{\Lambda\Lambda'} \, f_1^g(x,\bfk^2) 
+ \frac{(\epsilon^{{\bf \sigma}\bfk})_{\Lambda\Lambda'}}{M_N} \, 
f_{1T}^{\perp g}(x,\bfk^2) \Big) \nonumber\\
&-& i \epsilon^{ij}_T \,   
  \Big(\sigma^3_{\Lambda\Lambda'} \, g_{1L}^g(x,\bfk^2) 
+ \frac{({\bf\sigma} \bfk)_{\Lambda\Lambda'}}{M_N}  
\, g_{1T}^g(x,\bfk^2) \Big) \nonumber\\
&+& \eta^{ij}_T \, 
  \Big(\delta_{\Lambda\Lambda'} \, h_1^{(1) \perp g}(x,\bfk^2) 
+ \frac{(\epsilon^{{\bf \sigma}\bfk})_{\Lambda\Lambda'}}{M_N} 
\, h_{1T}^{- g}(x,\bfk^2)\Big) 
\nonumber\\ 
&+& \xi^{ij}_T \, 
  \Big(\sigma^3_{\Lambda\Lambda'} \, h_{1L}^{(1) \perp g}(x,\bfk^2) 
+ \frac{({\bf\sigma} \bfk)_{\Lambda\Lambda'}}{M_N} \, h_{1T}^{+ g}(x,\bfk^2)\Big) 
\,,
\en  
where $\epsilon^{\sigma\bfk} = \epsilon^{mn} \, \sigma^m \, \bfk^n 
= \sigma^1 k^2 - \sigma^2 k^1$.  

Further, one can expand the gluon tensor in the basis of gluon polarizations, 
using the spin density matrix $\rho_{\lambda\lambda'}^{ij}$. Using 
\eq 
\Gamma^{ij}_{\Lambda\Lambda'}(x,\bfk) = \sum\limits_{\lambda,\lambda'}
\, \rho_{\lambda\lambda'}^{ij} \,
\Gamma_{\lambda\lambda';\Lambda\Lambda'}(x,\bfk) \,,
\en   
one gets 
\eq 
\Gamma^{ij}(x,\bfk,{\bf S}) = \sum\limits_{\Lambda,\Lambda'} 
\, \rho_{\Lambda'\Lambda}({\bf S}) \, \rho_{\lambda\lambda'}^{ij} \, 
\Gamma_{\lambda\lambda';\Lambda\Lambda'}(x,\bfk) \,. 
\en 
From Eqs.~(\ref{gmunu_matrix}), (\ref{emunu_matrix}), 
(\ref{eta_circ}), and~(\ref{xi_circ}), 
we get for the circular polarization of the gluon: 
\eq  
\Gamma_{\lambda\lambda';\Lambda\Lambda'}(x,\bfk) &=& 
\delta_{\lambda\lambda'} \, 
\Big(\delta_{\Lambda\Lambda'} \, f_1^g(x,\bfk^2) 
+ \frac{(\epsilon^{{\bf \sigma}\bfk})_{\Lambda\Lambda'}}{M_N} \, 
f_{1T}^{\perp g}(x,\bfk^2) \Big) \nonumber\\
&+& \tau^3_{\lambda\lambda'} \, 
  \Big(\sigma^3_{\Lambda\Lambda'} \, g_{1L}^g(x,\bfk^2) 
+ \frac{({\bf\sigma} \bfk)_{\Lambda\Lambda'}}{M_N}  
\, g_{1T}^g(x,\bfk^2) \Big) \nonumber\\
&+& \eta_{\lambda\lambda'} \, 
  \Big(\delta_{\Lambda\Lambda'} \, h_1^{(1) \perp g}(x,\bfk^2) 
+ \frac{(\epsilon^{{\bf \sigma}\bfk})_{\Lambda\Lambda'}}{M_N} 
\, h_{1T}^{- g}(x,\bfk^2)\Big) 
\nonumber\\ 
&+& \xi_{\lambda\lambda'} \, 
  \Big(\sigma^3_{\Lambda\Lambda'} \, h_{1L}^{(1) \perp g}(x,\bfk^2) 
+ \frac{({\bf\sigma} \bfk)_{\Lambda\Lambda'}}{M_N} \, h_{1T}^{+ g}(x,\bfk^2)\Big) 
\,,
\en  
Explicitly, the $4 \otimes 4$ matrix $\Gamma^{\lambda\lambda'}_{\Lambda\Lambda'}$ in the 
gluon-nucleon basis $|i, \Lambda\ra = |+, +\ra$, $|+, -\ra$, $|-, +\ra$, $|+, +\ra$, 
is given by~\cite{Mulders:2000sh}:   
\eq 
\hspace*{-.75cm}
\left( 
\begin{array}{cccc}
f_1^g + g_{1L}^g
& \frac{|\bfk|}{M_N} \, e^{-i\phi_k} \, 
\Big[g_{1T}^g + i f_{1T}^{\perp g}\Big] 
& - e^{-2i\phi_k} \, 
\Big[h_1^{(1) \perp g} - i h_{1L}^{(1) \perp g}\Big]
& \frac{i |\bfk|}{M_N} \, e^{-3i\phi_k} \, 
\Big[h_{1T}^{+ g} - h_{1T}^{- g}\Big]
 \\[3mm]
\frac{|\bfk|}{M_N} \, e^{i\phi_k} \, 
\Big[g_{1T}^g- i f_{1T}^{\perp g}\Big] 
& f_1^g - g_{1L}^g
& \frac{i |\bfk|}{M_N} \, e^{-i\phi_k} \, 
\Big[h_{1T}^{+ g} + h_{1T}^{- g}\Big]
& - e^{-2i\phi_k} \, 
\Big[h_1^{(1) \perp g} + i h_{1L}^{(1) \perp g}\Big]
\\[3mm]
- e^{2i\phi_k} \, 
\Big[h_1^{(1) \perp g} + i h_{1L}^{(1) \perp g}\Big]
& - \frac{i |\bfk|}{M_N} \, e^{i\phi_k} \, 
\Big[h_{1T}^{+ g} + h_{1T}^{- g}\Big]
& f_1^g - g_{1L}^g 
& - \frac{|\bfk|}{M_N} \, e^{-i\phi_k} \, 
\Big[g_{1T}^g - i f_{1T}^{\perp g}\Big] 
\\[3mm]
- \frac{i |\bfk|}{M_N} \, e^{3i\phi_k} \, 
\Big[h_{1T}^{+ g} - h_{1T}^{- g}\Big]
& - e^{2i\phi_k} \, 
\Big[h_1^{(1) \perp g} - i h_{1L}^{(1) \perp g}\Big]
& - \frac{|\bfk|}{M_N} \, e^{i\phi_k} \, 
\Big[g_{1T}^g + i f_{1T}^{\perp g}\Big] 
& f_1^g + g_{1L}^g 
\end{array} 
\right) 
\en 
or in more compact form: 
\eq 
\hspace*{-.75cm}
\left( 
\begin{array}{cccc}
F_1^+                           & F_{1T}^+                 & H_1^-                      & \Delta H_{1T} \\
[1mm]
\Big(F_{1T}^+\Big)^\dagger      & F_1^-                    & H_{1T}                     & H_1^+         \\
[1mm]
\Big(H_1^-\Big)^\dagger         & \Big(H_{1T}\Big)^\dagger & F_1^-                      & F_{1T}^-      \\
[1mm]
\Big(\Delta H_{1T}\Big)^\dagger & \Big(H_1^+\Big)^\dagger  & \Big(F_{1T}^-\Big)^\dagger & F_1^+         \\
\end{array} 
\right) \,,
\en 
where 
\eq 
& &
F_1^\pm = f_1^g \pm g_{1L}^g\,, \qquad 
F_{1T}^\pm = \pm \frac{|\bfk|}{M_N} \, e^{- i \phi_k} \, 
\Big[g_{1T}^g \pm i f_{1T}^{\perp g}\Big] \,, \qquad 
H_1^\pm = - e^{-2i\phi_k} \, 
\Big[h_1^{(1) \perp g} \pm i h_{1L}^{(1) \perp g}\Big]\,, 
\nonumber\\
& & 
H_{1T} = \frac{i |\bfk|}{M_N} \, e^{-i\phi_k} \, 
\Big[h_{1T}^{+ g} + h_{1T}^{- g}\Big]\,, \quad 
\Delta H_{1T} = \frac{i |\bfk|}{M_N} \, e^{-3i\phi_k} \, 
\Big[h_{1T}^{+ g} - h_{1T}^{- g}\Big] \,.
\en 

At small $x$, the gluon TMDs are studied in Refs.~\cite{Boer:2015pni,Boer:2016xqr}. 
In the case of unpolarized tensor, it was found that~\cite{Boer:2016xqr} 
\eq 
\frac{x}{2} \Gamma_U^{ij}(x,\bfk) \, \stackrel{x \to 0}{\longrightarrow} \, 
\frac{\bfk^i \bfk^j}{2 M_N^2} \, e_U(\bfk^2) 
= \frac{1}{2} \, \biggl[- g_T^{ij} \, \frac{\bfk^2}{2 M_N^2} \, e_U(\bfk^2) 
\, + \, \eta_T^{ij} \, \frac{\bfk^2}{2 M_N^2} \, e_U(\bfk^2) 
\biggr],
\en 
which led to the identity~\cite{Boer:2016xqr} 
\eq
\lim\limits_{x \to 0} \, x f_1^g(x,\bfk^2) = 
\lim\limits_{x \to 0} \, x h_1^{(1) \perp g}(x,\bfk^2) = \frac{\bfk^2}{2 M_N^2} 
\, e_U(\bfk^2) = e_U^{(1)}(\bfk^2) 
\,, 
\en 
where $e_U(\bfk^2)$ is the scalar function defining the unpolarized gluon Wilson 
loop light-front correlator at small $x$ (see the exact expression in Eq.~(B.23) of 
Ref.~\cite{Boer:2016xqr}). 

Taking into account the positivity bounds, it follows that at small $x$ 
the Boer-Mulders TMD $h_1^{\perp g}$ becomes maximal 
$h_1^{\perp g}(x,\bfk^2) = (2 M_N^2/\bfk^2) \, f_1^g(x,\bfk^2)$. 
In Ref.~\cite{Boer:2015pni}, 
a relation between three the T-odd TMDs is found
\eq\label{identity_x0} 
x f_{1T}^{\perp g}(x,\bfk^2)  
= x h_{1T}^g(x,\bfk^2)   
= x h_{1T}^{\perp g}(x,\bfk^2)   
= - \frac{N_c}{4 \pi^2 \alpha_s} \, \bfk^2 \, O_{1T}^\perp(\bfk^2) \,, 
\en 
where $O_{1T}^\perp(\bfk^2)$ is the Odderon operator, 
consistent with the result of Ref.~\cite{Boer:2016xqr}, 
$N_c$ is the number of colors, and $\alpha_s$ is the strong coupling 
constant. Note the Odderon scales as $\alpha_s^3$~\cite{Zhou:2013gsa}.  
From Eq.~(\ref{identity_x0}), it follows that the vanishing of the transverse T-odd TMDs 
in the $\alpha_s \to 0$ limit occurs since they are proportional to $\alpha_s^2$. 

In particular, in Ref.~\cite{Boer:2016xqr},  
an expression for the tensor $\Gamma^{ij}_T(x,\bfk)$ 
at $x \to 0$ is deduced,  
\eq\label{GT_smallx} 
\frac{x}{2} \Gamma_T^{ij}(x,\bfk) \, \stackrel{x \to 0}{\longrightarrow} \, 
\frac{\bfk^i \bfk^j}{2 M_N^2} \, 
\frac{\epsilon_T^{{\bf S}_T\bfk}}{M_N} 
\, e_T(\bfk^2) = \frac{\epsilon_T^{{\bf S}_T\bfk}}{2 M_N} \, \biggl[ 
- g_T^{ij}    \, \frac{\bfk^2}{2 M_N^2} \, e_T(\bfk^2)
\, + \, 
  \eta_T^{ij} \, \frac{\bfk^2}{2 M_N^2} \, e_T(\bfk^2) 
\biggr] \,, 
\en 
where $e_T(\bfk^2)$ is the scalar function defining transverse part of the  gluon Wilson 
loop light-front correlator at small $x$ (see the exact expression in Eq.~(B.24) of 
Ref.~\cite{Boer:2016xqr}). 

From Eq.~(\ref{GT_smallx}), relations between TMDs follow
(here, we present TMDs used in our paper and in Ref.~\cite{Boer:2016xqr}):  
\eq 
  \lim\limits_{x \to 0} \, x f_{1T}^{\perp g}(x,\bfk^2) 
&=& \lim\limits_{x \to 0} \, x h_{1T}^{- g}(x,\bfk^2) 
= \lim\limits_{x \to 0} \, x h_1(x,\bfk^2) 
= - \frac{\bfk^2}{2 M_N^2} \, \lim\limits_{x \to 0} \, x h_{1T}^{\perp}(x,\bfk^2) 
\nonumber\\
&=& \frac{1}{2} \, \lim\limits_{x \to 0} \, x h_{1T}(x,\bfk^2) 
= \frac{\bfk^2}{2 M_N^2} \, e_T(\bfk^2) = e_T^{(1)}(\bfk^2) 
\en 
and 
\eq\label{h1Tplus_smallX} 
\lim\limits_{x \to 0} \, x h_{1T}^{+ g}(x,\bfk^2) = 0 \,. 
\en 
From Eq.~(\ref{h1Tplus_smallX}) it follows that the symmetric 
nucleon spin-gluon transverse momentum TMD $x h_{1T}^{+ g}(x,\bfk^2)$ 
vanishes at small $x$. 

It is also obtained in Refs.~\cite{Boer:2015pni,Boer:2016xqr} 
that the T worm-gear TMD $x g_{1T}^g(x,\bfk^2)$ vanishes at $x \to 0$. 
Vanishing of the longitudinal tensor $\frac{x}{2} \Gamma_T^{ij}(x,\bfk)$ at 
$x \to 0$ is proved in Ref.~\cite{Boer:2016xqr},   
\eq\label{GL_smallx} 
\frac{x}{2} \Gamma_L^{ij}(x,\bfk) = 0 
\en 
leading to the vanishing of the corresponding TMDs: 
\eq 
\lim\limits_{x \to 0} \, x g_{1L}^g(x,\bfk^2) = 0\,, 
\qquad  
\lim\limits_{x \to 0} \, x h_{1L}^{\perp g}(x,\bfk^2) = 0 
\,. 
\en 

In the next section, we will discuss results for the leading-order gluon TMDs 
in light-front (LF) QCD. In particular, we will show that LF QCD is consistent 
with model-independent results for the small-$x$ behavior of gluon TMDs, 
obtained in Refs.~\cite{Boer:2015pni,Boer:2016xqr}. Moreover, we will
discuss results of LF QCD for the large-$x$ scaling of gluon TMDs and a 
SR relating the T-even TMDs $f_1^g$, $g_{1L}^g$, $g_{1T}^g$,  
and $h_1^{\perp g}$. Based on this, we conjecture 
that it could follow from the condition 
${\rm det}\Big[\Gamma_{\lambda\lambda';\Lambda\Lambda'}(x,\bfk)\Big] = 0$.
 
\section{Gluon TMDs in LF QCD} 

In Ref.~\cite{Lyubovitskij:2020xqj}, 
following Refs.~\cite{Brodsky:2000ii,Lu:2016vqu}, 
we formulated a light-front model for the nucleon, motivated by soft-wall AdS/QCD, 
as a composite system of a spin-1 gluon and a spin-$\frac{1}{2}$ 
three-quark spectator system. The leading three-quark nucleon Fock state 
as a bound state of quark and diquark spectator was considered 
in Ref.~\cite{Lyubovitskij:2020otz}. 
In particular, we derived~\cite{Lyubovitskij:2020xqj} expressions 
for the T-even gluon TMDs at leading twist. Here, we extend 
our formalism into the T-odd gluon TMDs. 
First, we introduce the notation for the light-front wave function (LFWF) 
$\psi^{\Lambda}_{\lambda\lambda_X}(x,\bfk)$, consistent with the
previous section definitions and describing the
bound state of a gluon $(g)$ and a three-quark spectator $(X)$, 
where $\Lambda$, $\lambda$, and $\lambda_X$ are the helicities 
of nucleon, gluon, and three-quark spectator, respectively. 
For a nucleon with up helicity $\Lambda = + \frac{1}{2}$ (we use shorten notation $+$), 
four possible LFWFs are available, taking into account the different helicities 
$\lambda = \pm 1$ and $\lambda_X = \pm \frac{1}{2}$:  
\eq 
\psi^+_{+1 +\frac{1}{2}}(x,\bfk) &=& 
\frac{k^1-ik^2}{M_N} \, \varphi^{(2)}(x,\bfk^2) 
\,, 
\nonumber\\
\psi^+_{+1 -\frac{1}{2}}(x,\bfk) &=& \varphi^{(1)}(x,\bfk^2) \,, 
\nonumber\\
\psi^+_{-1 +\frac{1}{2}}(x,\bfk) &=& - \frac{k^1+ik^2}{M_N} \, (1-x) 
\, \varphi^{(2)}(x,\bfk^2) 
\,, 
\nonumber\\
\psi^+_{-1 -\frac{1}{2}}(x,\bfk) &=& 0 \,. 
\en 
The LFWFs with down helicity $\Lambda = - \frac{1}{2}$ (we use shorten notation $-$) 
are \eq\label{LFWF_in} 
\psi^-_{+1 +\frac{1}{2}}(x,\bfk) &=& 0 \,, \nonumber\\
\psi^-_{+1 -\frac{1}{2}}(x,\bfk) &=& 
- \Big[\psi^+_{-1 +\frac{1}{2}}(x,\bfk)\Big]^* =
  \frac{k^1-ik^2}{M_N} \, (1-x) \, \varphi^{(2)}(x,\bfk) 
\,, 
\nonumber\\
\psi^-_{-1 +\frac{1}{2}}(x,\bfk) &=& 
\Big[\psi^+_{+1 -\frac{1}{2}}(x,\bfk)\Big]^* = 
\varphi^{(1)}(x,\bfk^2) \,, 
\nonumber\\
\psi^-_{-1 -\frac{1}{2}}(x,\bfk) &=& 
- \Big[\psi^+_{+1 +\frac{1}{2}}(x,\bfk)\Big]^* = 
- \frac{k^1+ik^2}{M_N} \, \varphi^{(2)}(x,\bfk^2) 
\,, 
\en 
where the functions $\varphi^{(1)}(x,\bfk)$ and 
$\varphi^{(2)}(x,\bfk)$ can be expressed through the gluon PDFs 
\eq
G^\pm(x) = \frac{G(x) \pm \Delta G(x)}{2} \,, 
\en 
where $G(x)$ and $\Delta G(x)$ are the unpolarized 
and polarized gluon PDFs, as 
\eq\label{LFWFs_symmetric}
\varphi^{(1)}(x,\bfk^2) &=& 
\frac{4 \pi}{\kappa} \, \sqrt{G^+(x)} \, \beta(x) \, 
\sqrt{D_g(x)}  \, 
\exp\biggl[- \frac{\bfk^2}{2 \kappa^2} \, D_g(x) \biggr] \,,
\nonumber\\
\frac{1}{M_N} \,
\varphi^{(2)}(x,\bfk^2) &=& \frac{4 \pi}{\kappa^2} \, 
\sqrt{G^-(x)} \, \frac{D_g(x)}{1-x} \,
\exp\biggl[- \frac{\bfk^2}{2 \kappa^2} \, D_g(x) \biggr] \,, \\
\beta(x) &=& \sqrt{1 - \frac{G^-(x)}{G^+(x) (1-x)^2}} \,  
\,. \nonumber
\en 
Here, $D_g(x)$ is the profile function introduced 
in Refs.~\cite{Lyubovitskij:2020otz,Lyubovitskij:2020xqj}, 
modeling the $x$ dependence of the scale parameter: 
$\kappa^2(x) = \kappa^2/D_g(x)$. 
$D_g(x)$ is related to the function $f_g(x)$ as~\cite{Lyubovitskij:2020otz} 
\eq
D_g(x) = - \frac{\log[1- [f_g(x)]^{2/5} (1-x)^2]}{(1-x)^2} \,.
\en
On the other hand, the function $f_g(x)$ is fixed from the gluon PDF $G(x)$ 
by solving differential equation with respect to the $x$ variable: 
\eq
\biggl[ - f_g(x) (1-x)^5 \biggr]' = x G(x) \,. 
\en 
Note that in Ref.~\cite{Lyubovitskij:2020xqj}  
a generalization to the case in which the LFWFs 
$\varphi^{(1)}(x,\bfk^2)$ and $\varphi^{(1)}(x,\bfk^2)$ 
are expressed in terms of two different functions $D_{g_1}(x)$
and $D_{g_2}(x)$, respectively, is discussed in detail. 
There it is shown that the {\it Generalized version} fulfills all model-independent 
constraints as soon as the latter hold without reference to 
a specific choice of LFWFs. 
One can see that the $\varphi^{(2)}(x,\bfk^2)$ function is expressed 
in terms of the $G^-(x)$ PDF, while the 
$\varphi^{(1)}(x,\bfk^2)$ is expressed through the combination of the
functions $G^+(x)$ and $G^-(x)$. Notice that the ansatz~(\ref{LFWFs_symmetric}) 
for the LFWF is consistent with QCD results for gluon PDFs at 
large and small $x$, derived in Refs.~\cite{Brodsky:1989db,Brodsky:1994kg}. 
However, our results are not restricted to specific parametrizations of the gluon PDFs.
In fact, all our results for the gluon parton densities are expressed in terms
of their PDFs $G(x)$ and $\Delta G(x)$, which should be fixed from data. 
In particular, in Ref.~\cite{Avakian:2007xa}, 
it was shown that the large $x$ behavior of the quark PDFs  is modified
by large logarithmic corrections induced by the quark nonzero ($L_z \neq 0)$.
The prediction of Ref.~\cite{Avakian:2007xa} was motivated by predating experiments 
by the JLab Hall A Collaboration~\cite{Zheng:2003un} and 
by the CLAS Collaboration~\cite{Dharmawardane:2006zd}  
and later was experimentally confirmed 
by the JLab Hall A Collaboration~\cite{Parno:2014xzb}. 
One can expect that a similar modification could occur in the case of
gluon TMDs, since gluons are also expected to carry orbital angular momentum.
If such modification occurs, it does not change our results and predictions for the
parametrization of the gluon TMDs, while the choice of their PDFs as input quantities 
should be consistent with possible modification of their large $x$ behavior. 

Now we are in the position to derive light-front formalism expressions for all gluon TMDs at 
leading twist, i.e., including T-odd TMDs. 
The gluon correlator $\Gamma_{\lambda\lambda';\Lambda\Lambda'}(x,\bfk)$ 
in LF QCD reads:  
\eq 
\Gamma_{\lambda\lambda';\Lambda\Lambda'}(x,\bfk) = 
\sum\limits_{\lambda_1\lambda_2;\Lambda_1\Lambda_2;\lambda_X\lambda_X'} \, 
\psi^{* \Lambda_1}_{\lambda_1\lambda_X}(x,\bfk) \ 
G^{\lambda_1\lambda_2\Lambda_1\Lambda_2\lambda_X\lambda_X'}_{\lambda,\lambda';\Lambda\Lambda'}(x,\bfk) \ 
\psi^{\Lambda_2}_{\lambda_2\lambda_X'}(x,\bfk) \,, 
\en 
where $\psi^{\lambda_N}_{\lambda_g\lambda_X}$ is the gluon-three-quark LFWF, 
$G^{\lambda_1\lambda_2\Lambda_1\Lambda_2\lambda_X\lambda_X'}_{\lambda,\lambda';\Lambda\Lambda'}$ 
is the interaction kernel, which includes both T-even and T-odd (due to gluon rescattering
between gluon and three-quark (3q) spectator) structures. At leading twist, 
$G^{\lambda_1\lambda_2\Lambda_1\Lambda_2\lambda_X\lambda_X'}_{\lambda,\lambda';\Lambda\Lambda'}$ 
is decomposed into the eight independent tensorial structures, 
\eq 
G^{\lambda_1\lambda_2\Lambda_1\Lambda_2\lambda_X\lambda_X'}_{\lambda,\lambda';\Lambda\Lambda'}(x,\bfk) 
&=& 
\sum\limits_{i=1}^8 \, 
G^{\lambda_1\lambda_2\Lambda_1\Lambda_2\lambda_X\lambda_X'; (i)}_{\lambda,\lambda';\Lambda\Lambda'}(x,\bfk) 
\,, \nonumber\\
G^{\lambda_1\lambda_2\Lambda_1\Lambda_2\lambda_X\lambda_X'; (1)}_{\lambda,\lambda';\Lambda\Lambda'}(x,\bfk) 
&=& 
\frac{1}{32 \pi^3} \, 
\delta_{\lambda\lambda'} \, 
\delta_{\Lambda\Lambda'} \, 
\delta_{\lambda_1\lambda_2} \, 
\delta_{\Lambda_2\Lambda_1} \, 
\delta_{\lambda_X\lambda_X'} \,, \nonumber\\
G^{\lambda_1\lambda_2\Lambda_1\Lambda_2\lambda_X\lambda_X'; (2)}_{\lambda,\lambda';\Lambda\Lambda'}(x,\bfk) 
&=& 
\frac{1}{32 \pi^3} \, 
\tau^3_{\lambda\lambda'} \, 
\sigma^3_{\Lambda\Lambda'} \, 
\tau^3_{\lambda_1\lambda_2} \, 
\sigma^3_{\Lambda_2\Lambda_1} \, 
\delta_{\lambda_X\lambda_X'} \,, \nonumber\\
G^{\lambda_1\lambda_2\Lambda_1\Lambda_2\lambda_X\lambda_X'; (3)}_{\lambda,\lambda';\Lambda\Lambda'}(x,\bfk) 
&=& 
\frac{1}{32 \pi^3} \, 
\tau^3_{\lambda\lambda'} \, 
\frac{({\bf\sigma} \bfk)_{\Lambda\Lambda'}}{M_N}  \, 
\tau^3_{\lambda_1\lambda_2} \, 
\frac{({\bf\sigma} \bfk)_{\Lambda_2\Lambda'}}{M_N} \, 
\delta_{\lambda_X\lambda_X'} \,, \nonumber\\ 
G^{\lambda_1\lambda_2\Lambda_1\Lambda_2\lambda_X\lambda_X'; (4)}_{\lambda,\lambda';\Lambda\Lambda'}(x,\bfk) 
&=& 
\frac{1}{32 \pi^3} \, 
\eta_{\lambda\lambda'} \, 
\delta_{\Lambda\Lambda'} \, 
\eta_{\lambda_1\lambda_2} \,
\delta_{\Lambda_2\Lambda_1} \,  
\delta_{\lambda_X\lambda_X'} \,, \nonumber\\
G^{\lambda_1\lambda_2\Lambda_1\Lambda_2\lambda_X\lambda_X'; (5)}_{\lambda,\lambda';\Lambda\Lambda'}(x,\bfk) 
&=& 
\frac{1}{32 \pi^3} \, 
\xi_{\lambda\lambda'} \, 
\sigma^3_{\Lambda\Lambda'} \, 
(\tau^3\xi)_{\lambda_1\lambda_2} \,
\sigma^3_{\Lambda_2\Lambda_1} \,  
\tau^3_{\lambda_X\lambda_X'} \, i R_{h_{1L}^g}(x,\bfk^2) 
\,, \nonumber\\
G^{\lambda_1\lambda_2\Lambda_1\Lambda_2\lambda_X\lambda_X'; (6)}_{\lambda,\lambda';\Lambda\Lambda'}(x,\bfk) 
&=& 
\frac{1}{32 \pi^3} \, 
\xi_{\lambda\lambda'} \, 
\frac{({\bf\sigma} \bfk)_{\Lambda\Lambda'}}{M_N}  \,  
(\xi\tau^3)_{\lambda_1\lambda_2} \, 
\frac{({\bf\sigma} \bfk)_{\Lambda_2\Lambda_1}}{M_N}  \,  
\tau^3_{\lambda_X\lambda_X'} \, i R_{h_{1T}^{+g}}(x,\bfk^2) 
\,, \nonumber\\
G^{\lambda_1\lambda_2\Lambda_1\Lambda_2\lambda_X\lambda_X'; (7)}_{\lambda,\lambda';\Lambda\Lambda'}(x,\bfk) 
&=& 
\frac{1}{32 \pi^3} \, 
\eta_{\lambda\lambda'} \, 
\frac{(\epsilon^{{\bf \sigma}\bfk})_{\Lambda\Lambda'}}{M_N} 
\eta_{\lambda_1\lambda_2} \, 
\frac{(\epsilon^{{\bf \sigma}\bfk} \sigma^3)_{\Lambda_2\Lambda_1}}{M_N} 
\tau^3_{\lambda_X\lambda_X'} \, i R_{h_{1T}^{-g}}(x,\bfk^2) 
\,, \nonumber\\
G^{\lambda_1\lambda_2\Lambda_1\Lambda_2\lambda_X\lambda_X'; (8)}_{\lambda,\lambda';\Lambda\Lambda'}(x,\bfk) 
&=& 
\frac{1}{32 \pi^3} \, 
\delta_{\lambda\lambda'} \, 
\frac{(\epsilon^{{\bf \sigma}\bfk})_{\Lambda\Lambda'}}{M_N} 
\delta_{\lambda_1\lambda_2} \, 
\frac{(\epsilon^{{\bf \sigma}\bfk} \sigma^3)_{\Lambda_2\Lambda_1}}{M_N} 
\tau^3_{\lambda_X\lambda_X'} \, i R_{f_{1T}^{\perp g}}(x,\bfk^2) 
\,, 
\en 
which accordingly generate eight leading-twist gluon TMDs,  
\eq 
\delta_{\lambda\lambda'} \, 
\delta_{\Lambda\Lambda'} \, 
f_1^g(x,\bfk^2) &=& 
\Gamma_{\lambda\lambda';\Lambda\Lambda'}^{(1)}(x,\bfk) 
\,, \nonumber\\
\tau^3_{\lambda\lambda'} \, 
\sigma^3_{\Lambda\Lambda'} \, g_{1L}^g(x,\bfk^2) &=& 
\Gamma_{\lambda\lambda';\Lambda\Lambda'}^{(2)}(x,\bfk) 
\,, \nonumber\\
\tau^3_{\lambda\lambda'} \, 
\frac{({\bf\sigma} \bfk)_{\Lambda\Lambda'}}{M_N}  \, 
g_{1T}^g(x,\bfk^2) &=& 
\Gamma_{\lambda\lambda';\Lambda\Lambda'}^{(3)}(x,\bfk) 
\,, \nonumber\\
\eta_{\lambda\lambda'} \, 
\delta_{\Lambda\Lambda'} \, 
h_1^{(1) \perp g}(x,\bfk^2) &=& 
\Gamma_{\lambda\lambda';\Lambda\Lambda'}^{(4)}(x,\bfk) 
\,, \nonumber\\
\xi_{\lambda\lambda'} \, 
\sigma^3_{\Lambda\Lambda'} \, 
h_{1L}^{(1) \perp g}(x,\bfk^2) &=& 
\Gamma_{\lambda\lambda';\Lambda\Lambda'}^{(5)}(x,\bfk) 
\,, \nonumber\\
\xi_{\lambda\lambda'} \, 
\frac{({\bf\sigma} \bfk)_{\Lambda\Lambda'}}{M_N}  \, 
h_{1T}^{+g}(x,\bfk^2) &=& 
\Gamma_{\lambda\lambda';\Lambda\Lambda'}^{(6)}(x,\bfk) 
\,, \nonumber\\
\eta_{\lambda\lambda'} \, 
\frac{(\epsilon^{{\bf \sigma}\bfk})_{\Lambda\Lambda'}}{M_N} \, 
h_{1T}^{-g}(x,\bfk^2) &=& 
\Gamma_{\lambda\lambda';\Lambda\Lambda'}^{(7)}(x,\bfk) 
\,, \nonumber\\
\delta_{\lambda\lambda'}  \, 
\frac{(\epsilon^{{\bf \sigma}\bfk})_{\Lambda\Lambda'}}{M_N} 
f_{1T}^g(x,\bfk^2) &=& 
\Gamma_{\lambda\lambda';\Lambda\Lambda'}^{(8)}(x,\bfk) \,, 
\en 
where 
\eq 
\Gamma_{\lambda\lambda';\Lambda\Lambda'}(x,\bfk) 
&=& \sum\limits_{i=1}^8 \ \Gamma_{\lambda\lambda';\Lambda\Lambda'}^{(i)}(x,\bfk) \,, 
\nonumber\\
\Gamma_{\lambda\lambda';\Lambda\Lambda'}^{(i)}(x,\bfk) &=& 
\sum\limits_{\lambda_1\lambda_2;\Lambda_1\Lambda_2;\lambda_X\lambda_X'} \, 
\psi^{* \Lambda_1}_{\lambda_1\lambda_X}(x,\bfk) \ 
G^{\lambda_1\lambda_2\Lambda_1\Lambda_2\lambda_X\lambda_X'; (i)}_{\lambda,\lambda';\Lambda\Lambda'}(x,\bfk) \ 
\psi^{\Lambda_2}_{\lambda_2\lambda_X'}(x,\bfk) \,. 
\en 
Here, the loop functions 
$R_{h_{1L}^{(1) \perp g}}(x,\bfk^2)$ 
$R_{f_{1T}^{\perp g}}(x,\bfk^2)$, and 
$R_{h_{1T}^{\pm g}}(x,\bfk^2)$ encode rescattering between the gluon 
and the three-quark spectator system. 

Analytical expressions for the gluon TMDs in terms of LFWFs are: 
\eq 
f_1^{g}(x,\bfk^2) &=& 
\frac{1}{16 \pi^3} \, \biggl[
\Big(\varphi^{(1)}(x,\bfk^2)\Big)^2
+ \frac{\bfk^2}{M_N^2} \, \Big[1+(1-x)^2\Big]
\, \Big(\varphi^{(2)}(x,\bfk^2)\Big)^2 \biggr] \,, \nonumber\\
g_{1L}^{g}(x,\bfk^2) &=& 
\frac{1}{16 \pi^3} \, \biggl[
\Big(\varphi^{(1)}(x,\bfk^2)\Big)^2
+ \frac{\bfk^2}{M_N^2} \, \Big[1-(1-x)^2\Big]
\, \Big(\varphi^{(2)}(x,\bfk^2)\Big)^2 \biggr] \,, \nonumber\\
g_{1T}^{g}(x,\bfk^2) &=& 
\frac{1}{8 \pi^3} \,
\varphi^{(1)}(x,\bfk^2) \, \varphi^{(2)}(x,\bfk^2) \, (1-x) \,, \nonumber\\
h_1^{(1) \perp g}(x,\bfk^2) &=& 
\frac{1}{8 \pi^3} \, \frac{\bfk^2}{M_N^2} \, 
\Big[\varphi^{(2)}(x,\bfk^2)\Big]^2 \, (1-x) \, \label{f1g}
\en 
and 
\eq 
h_{1L}^{(1) \perp g}(x,\bfk^2) &=& 
\frac{1}{8 \pi^3} \, \frac{\bfk^2}{M_N^2} \, 
\Big[\varphi^{(2)}(x,\bfk^2)\Big]^2 \, (1-x) 
\, R_{h_{1L}^{(1) \perp g}}(x,\bfk^2)
\,, \nonumber\\
h_{1T}^{\pm g}(x,\bfk^2) &=& 
\frac{1}{8 \pi^3} \, 
\varphi^{(1)}(x,\bfk^2) \, \varphi^{(2)}(x,\bfk^2) \, 
R_{h_{1T}^{\pm g}}(x,\bfk^2) 
\,, \nonumber\\
f_{1T}^{\perp g}(x,\bfk^2) &=& \frac{1}{8 \pi^3}
\, \varphi^{(1)}(x,\bfk^2) \, \varphi^{(2)}(x,\bfk^2) \,
(1-x) \,  R_{f_{1T}^{\perp g}}(x,\bfk^2) \,. \label{h1Lg}
\en 
Without referring to the specific choice of the LFWFs $\varphi^{(1,2)}$,  
we can express the T-odd TMDs in terms of T-even ones: 
\eq 
h_{1L}^{(1) \perp g} &=& \frac{f_1^g(x,\bfk^2) - g_{1L}^g(x,\bfk^2)}{1-x} \, 
R_{h_{1L}^{(1)\perp g}}(x,\bfk^2) = 
h_1^{(1) \perp g}(x,\bfk^2) \, R_{h_{1L}^{(1)\perp g}}(x,\bfk^2) 
\,, \label{h1Lx0}\\
h_{1T}^{\pm g} &=& \frac{g_{1T}^g(x,\bfk^2)}{1-x} \, 
R_{h_{1T}^{\pm g}}(x,\bfk^2) \,, \label{h1Tx0}\\
f_{1T}^{\perp g} &=& 
g_{1T}^g(x,\bfk^2) \, R_{f_{1T}^{\perp g}}(x,\bfk^2) \,.  
\label{f1Tx0}
\en 
In Ref.~\cite{Lyubovitskij:2020xqj}, using parametrizations of 
the LFWFs~(\ref{LFWFs_symmetric}) 
in terms of gluon PDFs,  
we derived expressions for the T-even gluon TMDs: 
\eq
f_1^g(x,\bfk^2) &=& \frac{D_g(x)}{\pi \kappa^2} \,
\biggl[
G(x)  + G^-(x) \, \alpha_+(x) \, \biggl(
\frac{\bfk^2}{\kappa^2} \, D_g(x)  - 1 \biggr) \biggr] \,
\exp\biggl[- \frac{\bfk^2}{\kappa^2} \, D_g(x) \biggr] \,, \nonumber\\
g_{1L}^g(x,\bfk^2) &=& \frac{D_g(x)}{\pi \kappa^2} \,
\biggl[
\Delta G(x)  + G^-(x) \, \alpha_-(x) \, \biggl(
\frac{\bfk^2}{\kappa^2} \, D_g(x)  - 1 \biggr) \biggr] \,
\exp\biggl[- \frac{\bfk^2}{\kappa^2} \, D_g(x) \biggr] \,, \nonumber\\
g_{1T}^g(x,\bfk^2) &=& \frac{D_g^{3/2}(x) M_N}{\pi \kappa^3} \,
\sqrt{G^2(x)-\Delta G^2(x)} \ \beta(x)
\, \exp\biggl[- \frac{\bfk^2}{\kappa^2} \, D_g(x) \biggr] \,, \nonumber\\
h_1^{(1) \perp g}(x,\bfk^2) &=& \bfk^2 \, \frac{D_g^2(x)}{\pi \kappa^4}
\, \frac{G(x)-\Delta G(x)}{1-x} \
\, \exp\biggl[- \frac{\bfk^2}{\kappa^2} \, D_g(x) \biggr] \,, \label{f1g_QCD} 
\en
where 
\eq
\alpha_\pm(x) = \frac{1 \pm  (1-x)^2}{(1-x)^2} \,.
\en
For convenience, one can present T-even gluon TMDs in an equivalent form: 
\eq 
f_1^g(x,\bfk^2) &=& G(x) \, \frac{D_g(x)}{\pi \kappa^2} \,
\biggl[\sigma_+(x) + \frac{\bfk^2}{\kappa^2} \, D_g(x) \, [1-\sigma_+(x)] 
\biggr] \,
\exp\biggl[- \frac{\bfk^2}{\kappa^2} \, D_g(x) \biggr] \,, 
\nonumber\\
g_{1L}^g(x,\bfk^2) &=& \Delta G(x) \, \frac{D_g(x)}{\pi \kappa^2} \,
\biggl[\sigma_-(x) + \frac{\bfk^2}{\kappa^2} \, D_g(x) \, [1-\sigma_-(x)] 
\biggr] \, 
\exp\biggl[- \frac{\bfk^2}{\kappa^2} \, D_g(x) \biggr] \,, 
\nonumber\\
&=& G(x) \, \frac{D_g(x)}{\pi \kappa^2} \,
\biggl[\sigma_+(x) + \frac{\bfk^2}{\kappa^2} \, D_g(x) 
\, \rho(x) \, [1-\sigma_-(x)] 
\biggr] \,  
\exp\biggl[- \frac{\bfk^2}{\kappa^2} \, D_g(x) \biggr] \,, 
\nonumber\\
g_{1T}^g(x,\bfk^2) &=& G(x) \, \frac{D_g^{3/2}(x) M_N}{\pi \kappa^3} \,
\sqrt{2 \sigma_+(x) \, [1 - \rho(x)]} 
\, \exp\biggl[- \frac{\bfk^2}{\kappa^2} \, D_g(x) \biggr] \,, 
\nonumber\\ 
h_1^{(1) \perp g}(x,\bfk^2) &=& 
G(x) \, \bfk^2 \, \frac{D_g^2(x)}{\pi \kappa^4}
\, 
\sqrt{[1 - \rho(x)] \, [1 + \rho(x) - 2\sigma_+(x)]} 
\, \exp\biggl[- \frac{\bfk^2}{\kappa^2} \, D_g(x) \biggr] \,, 
\label{f1gG_QCD} 
\en 
where 
\eq\label{rho} 
\rho(x) = \frac{\sigma_+(x)}{\sigma_-(x)}\,, \qquad  
\sigma_\pm (x) = \frac{\beta^2(x)}{1 \pm (1-x)^2 (1-\beta^2(x))} \,.
\en 

\section{Sum rules for gluon TMDs} 

We start this section by stating the two important SRs, 
involving T-even TMDs derived in Ref.~\cite{Lyubovitskij:2020xqj}:   
\eq 
\Big[f_1^g(x,\bfk^2)\Big]^2 &=& 
\Big[g_{1L}^g(x,\bfk^2)\Big]^2 \,+\, 
\Big[g_{1T}^{(1/2) g}(x,\bfk^2)\Big]^2 \,+\, 
\Big[h_1^{(1) \perp g}(x,\bfk^2)\Big]^2 \,, 
\label{SR1}\\
f_1^g(x,\bfk^2) - g_{1L}^g(x,\bfk) &=& 
(1-x) \, h_1^{(1) \perp g}(x,\bfk^2) \,,  
\label{SR2} 
\en 
where 
\eq\label{TMD_even2} 
{\rm TMD}^{(1/2)}(x,\bfk^2) = \frac{|\bfk|}{M_N}  \, {\rm TMD}(x,\bfk^2) \,, 
\qquad 
{\rm TMD}^{(1)}(x,\bfk^2) = \frac{\bfk^2}{2M_N^2} \, {\rm TMD}(x,\bfk^2) \,.
\en  

Notice that these SRs follow directly from Eq.~(\ref{f1g}), without 
any referring to a specific choice of LFWFs. 
The first SR is very exciting because it establishes a relation between the square of 
unpolarized TMD [lhs of Eq.~(\ref{SR1})] and a superposition of squares of 
three polarized TMDs [rhs of Eq.~(\ref{SR1})]. From this SR, the MR inequalities 
(positivity bounds) involving the T-even gluon TMDs~\cite{Mulders:2000sh}, 
follow immediately: 
\eq\label{MR_bounds} 
& &
\sqrt{\Big[g_{1L}^g(x,\bfk^2)\Big]^2 \,+\, 
      \Big[g_{1T}^{(1/2) g}(x,\bfk^2)\Big]^2} 
\le f_1^g(x,\bfk^2) \,, \nonumber\\
& &
\sqrt{\Big[g_{1L}^g(x,\bfk^2)\Big]^2 \,+\, 
      \Big[h_1^{(1) \perp g}(x,\bfk^2)\Big]^2} 
\le f_1^g(x,\bfk^2) \,, \\
& &
\sqrt{\Big[g_{1T}^{(1/2) g}(x,\bfk^2)\Big]^2 \,+\,  
      \Big[h_1^{(1) \perp g}(x,\bfk^2)\Big]^2}
\le f_1^g(x,\bfk^2) \,. \nonumber 
\en 
Note that the MR inequalities, involving all eight TMDs, 
taking into account positivity bounds and positivity 
eigenvalues [see also Ref.~\cite{Bacchetta:1999kz} 
for the case of quark TMDs], read 
\eq 
& &\sqrt{\Big[g_{1L}^g(x,\bfk^2)\Big]^2 \,+\, 
      \Big[g_{1T}^{(1/2) g}(x,\bfk^2)\Big]^2 \,+\, 
      \Big[f_{1T}^{(1/2) \perp g}(x,\bfk^2)\Big]^2} 
\le f_1^g(x,\bfk^2) \,, \label{MR_1}\\[2mm]
& &\sqrt{\Big[g_{1L}^g(x,\bfk^2)\Big]^2 + 
      \Big[h_1^{(1) \perp g}(x,\bfk^2)\Big]^2 +  
      \Big[h_{1L}^{(1) \perp g}(x,\bfk^2)\Big]^2}  
\le f_1^g(x,\bfk^2) \,, \label{MR_2}\\[2mm]
& &\Big|h_{1T}^{(1/2) + g}(x,\bfk^2) \mp h_{1T}^{(1/2) + g}(x,\bfk^2)\Big| 
\le f_1^g(x,\bfk^2) \pm g_{1L}^g(x,\bfk^2) \,,  \label{MR_3}\\[2mm]
& &\sqrt{\Big[g_{1L}^{g}(x,\bfk^2) \mp h_{1T}^{(1/2) + g}(x,\bfk^2)\Big]^2 \,+\, 
         \Big[g_{1T}^{(1/2) g}(x,\bfk^2)  \pm h_{1L}^{(1) \perp g}(x,\bfk^2)\Big]^2 \,+\,          
         \Big[h_1^{(1) \perp g}(x,\bfk^2) \pm f_{1T}^{(1/2) \perp g}(x,\bfk^2)\Big]^2} 
\nonumber\\
& &\le f_1^g(x,\bfk^2) \pm h_{1T}^{(1/2) - g}(x,\bfk^2) \,.  \label{MR_4}
\en 
From the above inequalities, it follows that the seven gluon TMDs $H^g=g_{1L}^g$, 
$g_{1T}^{(1/2) g}$, $h_1^{(1) \perp g}$, $h_{1L}^{(1) \perp g}$, 
$f_{1T}^{(1/2) \perp g}$, $h_{1T}^{(1/2) \pm g}$ are bounded 
by the unpolarized TMD $f_1^g$: 
\eq 
|H^g(x,\bfk^2)| \le f_1^g(x,\bfk^2) \,.
\en 
 
Our SRs~(\ref{SR1}) and~(\ref{SR2}) are also consistent with the results 
obtained in Refs.~\cite{Boer:2015pni,Boer:2016xqr} 
for the small-$x$ behavior of gluon TMDs. 
In particular, according to Refs.~\cite{Boer:2015pni,Boer:2016xqr}, 
at small $x$, the TMDS $g_{1L}^g$ and 
$g_{1T}^{(1/2) g}$ vanish, and $f_1^g(x,\bfk^2) = h_1^{(1) \perp g}(x,\bfk^2)$. 
This is fully consistent with our SRs at small $x$. In particular, 
our SRs at small $x$ become degenerate and read 
\eq 
f_1^g(x,\bfk^2) = h_1^{(1) \perp g}(x,\bfk^2) \,. 
\en 
Note also that, using explicit expressions for our 
T-even TMDs, we confirm that the TMDS $g_{1L}^g$ and
$g_{1T}^{(1/2) g}$ vanish at small $x$.  
It also interesting to see what happens at large $x$ in LF QCD. 
In the limit of large $x$, the TMDs $g_{1T}^{(1/2) g}$ and 
$h_{1}^{(1) \perp g}$ vanish, and our two SRs~(\ref{SR1}) and~(\ref{SR2}) 
become degenerate, leading to the relation 
\eq
f_1^g(x,\bfk^2) = g_{1L}^g(x,\bfk) \,. 
\en 
Now we conjecture that the derived SR~(\ref{SR1}) 
is consequence of the condition 
\eq\label{detG}
{\rm det}\Big[\Gamma_{\lambda\lambda';\Lambda\Lambda'}\Big] = 0, 
\en 
signaling that the gluon TMDs are not independent and are
related via SRs. As soon as condition~(\ref{detG}) 
is expanded in powers of the strong coupling constant $\alpha_s$, from 
zero order up to the second order, it gives us 
three SRs at orders ${\cal O}(1)$, ${\cal O}(\alpha_s)$, 
and ${\cal O}(\alpha_s^2)$. In particular, from 
Eq.~(\ref{detG}), the condition 
\eq 
\Big[ R_0 + 2 R_1 + R_2 \Big] \, \Big[ R_0 - 2 R_1 + R_2 \Big] = 0 
\en 
follows, where $R_0$, $R_1$, and $R_2$ are the terms at orders  
${\cal O}(1)$, ${\cal O}(\alpha_s)$, 
and ${\cal O}(\alpha_s^2)$, respectively.    
Therefore, we have three SRs 
$R_0 = 0$, $R_1 = 0$, and $R_2 = 0$. 
Explicitly, $R_i$ are expressed in terms of gluon TMD as  
\eq 
R_0 &=& 
\Big[g_{1L}^g\Big]^2 \,+\, 
\Big[g_{1T}^{(1/2) g}\Big]^2 \,+\, 
\Big[h_1^{(1) \perp g}\Big]^2 - \Big[f_1^g\Big]^2  = 0\,, \label{SR1_v1}\\
R_1 &=&  f_1^g    \, h_{1T}^{(1/2) - g}
   \,+\, g_{1L}^g \, h_{1T}^{(1/2) + g}
   \,-\, g_{1T}^{(1/2) g}  \, h_{1L}^{(1) \perp g}
   \,-\, h_1^{(1) \perp g} \, f_{1T}^{(1/2) \perp g} = 0\,, \label{SR2_v1}\\
R_2 &=&  \Big[f_{1T}^{(1/2) \perp g}\Big]^2  
   \,+\, \Big[h_{1L}^{(1)    \perp g}\Big]^2  
   \,+\, \Big[h_{1T}^{(1/2) + g}\Big]^2 
   \,-\, \Big[h_{1T}^{(1/2) - g}\Big]^2 = 0 \label{SR3_v1} \,.
\en 
It is clear that the SR $R_0=0$ is exactly our first SR involving four 
T-even TMDs, the second SR $R_1=0$ couples T-even and T-odd TMDs, 
and finally the third SR $R_2=0$ involves only T-odd TMDs. 
We already stressed that the $R_0=0$ SR is consistent with the results of 
Refs.~\cite{Boer:2015pni,Boer:2016xqr}, which study gluon TMDs at small $x$. 
It is interesting that the other two SRs $R_1 = 0$ and $R_2 = 0$ are also consistent 
with the results of Refs.~\cite{Boer:2015pni,Boer:2016xqr}. 
In particular, at small $x$, we get: 
\eq 
& &f_1^g = h_1^{(1) \perp g}\,, \quad  h_{1T}^{(1/2) - g} =  f_{1T}^{(1/2) \perp g} \nonumber\\
& &g_{1L}^g = g_{1T}^g = h_{1T}^{(1/2) + g} = 0 \,. 
\en   
Finally, taking into account vanishing TMDs, the SRs $R_i = 0$  are simplified at small $x$ and 
read 
\eq 
R_0 &=&  \Big[h_1^{(1) \perp g}\Big]^2 \,-\, \Big[f_1^g\Big]^2  = 0\,, \\
R_1 &=&  f_1^g   \, h_{1T}^{(1/2) - g}
   \,-\, h_1^{(1) \perp g} \, f_{1T}^{(1/2) \perp g} = 0\,, \\
R_2 &=&  \Big[f_{1T}^{(1/2) \perp g}\Big]^2  
   \,-\, \Big|h_{1T}^{(1/2) - g}\Big]^2 = 0 \,.
\en 

We remind the reader that in Ref.~\cite{Lyubovitskij:2020xqj} we 
studied the large $x$ scaling of the T-even TMDs. We found that 
$f_1^g(x,\bfk^2)$ and $g_{1L}^g(x,\bfk^2)$ scale as $(1-x)^4$, 
while $g_{1T}^g(x,\bfk^2)$ and $h_1^{\perp g}(x,\bfk^2)$ have an
extra power $(1-x)$ falloff. Such large $x$ behavior of T-even TMDs 
is consistent with the first SR~(\ref{SR1}), which at large $x$ is simplified 
to 
\eq 
R_0 =
\Big[g_{1L}^g\Big]^2 \,-\, \Big[f_1^g\Big]^2  = 0 
\en 
or $|g_{1L}^g| = |f_1^g|$. 

Next, we take into account the following arguments at large $x$: 
(1) In the case of the T-even TMDs,  
$g_{1T}^{(1/2) g} \sim h_1^{(1) \perp g} \sim (1-x)^5$
are suppressed in comparison with $f_1^g \sim g_{1L}^g \sim (1-x)^4$~\cite{Lyubovitskij:2020xqj}; 
(2) In the case of the T-odd TMDs, the behavior of $f_{1T}^{(1/2) \perp g}$ and 
$h_{1L}^{(1) \perp g}$ is similar to the scaling 
of the $g_{1T}^{(1/2) g}$, up to corresponding loop factors 
$R_{h_{1L}^{(1)\perp g}}(x,\bfk^2)\bigg|_{x \to 1}$ and 
$R_{f_{1T}^{\perp g}}(x,\bfk^2)\bigg|_{x \to 1}$, which are expected to 
be constants or power of $(1-x)$.  

Therefore, at large $x$ the $R_1 = 0$,  
and $R_2 = 0$ are simplified as 
\eq 
R_1 &=&  f_1^g    \, \Big[h_{1T}^{(1/2) - g}
   \,+\, h_{1T}^{(1/2) + g} \Big] 
   = 0\,, \label{SR2_vhigh}\\
R_2 &=&  
         \Big[h_{1T}^{(1/2) + g}\Big]^2 
   \,-\, \Big[h_{1T}^{(1/2) - g}\Big]^2 = 0 \label{SR3_vhigh} \,.
\en 
From Eqs.~(\ref{SR2_vhigh}) and~(\ref{SR3_vhigh}), the 
constraint on $h_{1T}^{(1/2) \pm g}$ TMDs follows,  
\eq 
h_{1T}^{(1/2) + g} = - h_{1T}^{(1/2) - g} \,, 
\en 
which is consistent with the positivity bound~(\ref{MR_3}) at large $x$.
 
Now, let us consider the behavior of TMDs derived in LF QCD at small $x$. 
First, we look at the nucleon spin unpolarized TMDs $f_1^g$ and $h_1^{(1) \perp g}$,
where we reproduce the model-independent result 
of Ref.~\cite{Boer:2016xqr},  
\eq
\lim\limits_{x \to 0} \, x f_1^g(x,\bfk^2) = 
\lim\limits_{x \to 0} \, x h_1^{(1) \perp g}(x,\bfk^2) = 
\frac{\bfk^2}{2 M_N^2} \, e_U(\bfk^2) = e^{(1)}_U(\bfk^2) 
\,, 
\en 
where the function $e_U(\bfk^2)$ in case of LF QCD reads 
\eq 
e^{(1)}_U(\bfk^2) = \frac{\bfk^2}{\pi \kappa^4} \, D_g^2(0) \, 
\exp\Big(-\frac{\bfk^2}{\kappa^2} D_g(0)\Big) \, \lim\limits_{x \to 0} \, x G(x)  \,.
\en   
Integrating the function $e^{(1)}_U(\bfk^2)$ gives 
\eq 
\int d^2\bfk \, e^{(1)}_U(\bfk^2) = \lim\limits_{x \to 0} \, x G(x)  \,. 
\en 
Next, we look at the TMDs with longitudinal nucleon polarization 
$x g_{1L}^g$ and $x h_{1L}^{(1) \perp g}$, according to the model-independent consideration 
in Ref.~\cite{Boer:2016xqr}. In LF QCD, $x g_{1L}^g$ has an extra power of $x$ and 
also vanishes as in Ref.~\cite{Boer:2016xqr}:
\eq 
\lim\limits_{x \to 0} \, x g_{1L}^g(x,\bfk^2) = (N_q - 1) 
\frac{D_g(0)}{\pi \kappa^2} \, \biggl[ 1 + \frac{\bfk^2}{\kappa^2} \, \frac{D_g(0)}{N_q-1}\biggr]
\exp\Big(-\frac{\bfk^2}{\kappa^2} D_g(0)\Big) \, \lim\limits_{x \to 0} \, x^2 G(x) = 0 \,,
\en 
where $N_q$ is the number of valence quarks in a specific hadron (e.g.,
$N_q = 3$ in case of a nucleon). At small $x$, the gluon asymmetry ratio 
$g_{1L}^g(x,\bfk^2)/f_1^g(x,\bfk^2)$ behaves in LF QCD as 
\eq 
\frac{g_{1L}^g(x,\bfk^2)}{f_1^g(x,\bfk^2)}\bigg|_{x \to 0} = 
x \, \biggl[ 1 + \frac{\kappa^2}{\bfk^2} \, \frac{N_q-1}{D_g(0)} \biggr] \,. 
\en 
Notice that the gluon asymmetry ratio of integrated $g_{1L}^g$ and $f_1^g$ TMDs  
holds in the limit $x \to 0$,  
\eq\label{smallx}
\frac{\Delta G(x)}{G(x)} \to N_q x,   
\en 
which is consistent with Reggeon exchange arguments~\cite{Brodsky:1988ip,Brodsky:1989db}.

The small $x$ behavior of the $h_{1L}^{(1) \perp g}$ TMD can be 
understood using the relation~(\ref{h1Lx0}). In particular, 
we get 
\eq 
\frac{h_{1L}^{(1) \perp g}(x,\bfk^2)}{f_1^g(x,\bfk^2)}\bigg|_{x \to 0}  =  
R_{h_{1L}^{(1)\perp g}}(x,\bfk^2) \,.
\en 
Therefore, the vanishing of $x h_{1L}^g(x,\bfk^2)$ at $x \to 0$ 
means that the function $R_{h_{1L}^{(1)\perp g}}(x,\bfk^2)$ must also
vanish at $x \to 0$. 

Finally, we look at the TMDs with transverse nucleon polarization. 
First, we check what happens with the worm-gear TMD $x g_{1T}^g$ at small $x$ 
in LF QCD. At $x \to 0$, it has extra factor $\sqrt{x}$ in comparison 
with $x f_1^g$ and therefore is suppressed,  
\eq\label{g1Tg_f1g} 
\frac{g_{1T}^g(x,\bfk^2)}{f_1^g(x,\bfk^2)}\bigg|_{x \to 0}  =  
\frac{M_N \kappa}{\bfk^2 D_g^{1/2}(0)} \, \sqrt{x} \,,
\en 
which is again in agreement with result of Ref.~\cite{Boer:2016xqr}. 

Now, we are in a position to discuss the small $x$ behavior of the three 
T-odd transverse TMDs $f_{1T}^{\perp g}$, $h_{1T}^{+ g}$, and 
$h_{1T}^{- g}$. Because these functions contain unknown 
functions $R_{h_{1L}^{(1) \perp g}}(x,\bfk^2)$,  
$R_{f_{1T}^{\perp g}}(x,\bfk^2)$, and 
$R_{h_{1T}^{\pm g}}(x,\bfk^2)$ in LF QCD, encoding rescattering between the gluon 
and the three-quark spectator system, we can only get constraints on these functions 
at small $x$. In particular, using the relation of these functions with the $g_{1T}^g$ TMD 
[see Eqs.~(\ref{h1Tx0}) and~(\ref{f1Tx0})] and using Eq.~(\ref{g1Tg_f1g}), we get
\eq 
& &\frac{h_{1T}^{+ g}(x,\bfk^2)}{f_1^g(x,\bfk^2)}\bigg|_{x \to 0} 
\to \frac{M_N \kappa}{\bfk^2 D_g^{1/2}(0)} \, 
\Big[\sqrt{x} R_{h_{1T}^{+ g}}(x,\bfk^2)\Big]\bigg|_{x \to 0} = 0 \,, \\
& &\frac{h_{1T}^{- g}(x,\bfk^2)}{f_1^g(x,\bfk^2)}\bigg|_{x \to 0} 
\to \frac{M_N \kappa}{\bfk^2 D_g^{1/2}(0)} \, 
\Big[\sqrt{x} R_{h_{1T}^{- g}}(x,\bfk^2)\Big]\bigg|_{x \to 0} 
= \frac{e_T(\bfk^2)}{e_U(\bfk^2)} \,, \\
& &\frac{f_{1T}^{\perp g}(x,\bfk^2)}{f_1^g(x,\bfk^2)}\bigg|_{x \to 0} 
\to \frac{M_N \kappa}{\bfk^2 D_g^{1/2}(0)} \, 
\Big[\sqrt{x} R_{f_{1T}^{\perp g}}(x,\bfk^2)\Big]\bigg|_{x \to 0} 
= \frac{e_T(\bfk^2)}{e_U(\bfk^2)} \,. 
\en 
From the above equations, it is clear that the vanishing of $x h_{1T}^{+ g}(x,\bfk^2)$ 
at $x \to 0$ means that $R_{h_{1T}^{+ g}}(0,\bfk^2)$ is constant. 
The surviving of $x h_{1T}^{- g}(x,\bfk^2)$ and $x f_{1T}^{\perp g}(x,\bfk^2)$ 
at small $x$ leads to the constraints: 
\eq 
R_{h_{1T}^{- g}}(x,\bfk^2)\bigg|_{x \to 0} = 
R_{f_{1T}^{\perp g}}(x,\bfk^2)\bigg|_{x \to 0} =  
\sqrt{\frac{D_g(0)}{x}} \, \frac{\bfk^2}{M_N \kappa} \, 
\frac{e_T(\bfk^2)}{e_U(\bfk^2)} \,.  
\en 

\section{Conclusions}

In the present paper, we proposed a decomposition of 
the gluon correlator at leading twist, 
using as a basis four tensors (one antisymmetric 
and three symmetric). These tensors are expressed through 
generators of the $U(2)$ group acting in the 
two-dimensional plane of the transverse momentum of a gluon.
One of the new results is that all gluon TMDs stand as tensor structures, 
proportional to one of the basis set. We performed a clear 
interpretation of the two transversity T-odd TMDs with linear gluon
polarization: symmetric and asymmetric under permutation of 
the nucleon transverse spin and transverse momentum of gluon. 
Next, using LFWF representations, 
we derived T-even and T-odd gluon TMDs in the nucleon at leading twist.  
We construct the gluon-three-quark Fock component in the nucleon 
as a bound state of a gluon and a three-quark core. Gluon TMDs are given 
by factorized products of two LFWFs and a gluonic operator containing 
information about all TMDs at leading twist. We parametrize the effects 
of gluon rescattering in T-odd TMDs by four unknown scalar functions 
depending on two variables $x$ and $\bfk$. 
Derived gluon TDMs obey the model-independent MR 
inequalities~\cite{Mulders:2000sh}. Using our analytical 
results in LF QCD formalism,  we derived two new SRs involving T-even TMDs.  
One of these SRs states gives the identity between the square 
of the unpolarized TMD and the sum of the squares of three polarized TMDs. 
This SR gives more stringent limits on unpolarized TMDs, and 
all three MR inequalities between unpolarized and 
polarized TMDs directly follow from this SR. 
Based on this sum rule derived for the T-even TMDs or 
at order ${\cal O}(\alpha_s^0)$,  
we did a further conjecture, which involves all gluon TMDs in the condition 
${\rm det}\Big[\Gamma_{\lambda\lambda';\Lambda\Lambda'}(x,\bfk)\Big] = 0$ 
leads to three independent SRs at orders 
${\cal O}(\alpha_s^0)$, ${\cal O}(\alpha_s)$,
and ${\cal O}(\alpha_s^2)$.  

The SRs at arbitrary $x$ and $\bfk$ read
\eq 
R_0 &=& 
\Big[g_{1L}^g\Big]^2 \,+\, 
\Big[g_{1T}^{(1/2) g}\Big]^2 \,+\, 
\Big[h_1^{(1) \perp g}\Big]^2 - \Big[f_1^g\Big]^2  = 0\,, \label{SR1_sum}\\
R_1 &=&  f_1^g    \, h_{1T}^{(1/2) - g}
   \,+\, g_{1L}^g \, h_{1T}^{(1/2) + g}
   \,-\, g_{1T}^{(1/2) g}  \, h_{1L}^{(1) \perp g}
   \,-\, h_1^{(1) \perp g} \, f_{1T}^{(1/2) \perp g} = 0\,, \label{SR2_sum}\\
R_2 &=&  \Big[f_{1T}^{(1/2) \perp g}\Big]^2  
   \,+\, \Big[h_{1L}^{(1)    \perp g}\Big]^2  
   \,+\, \Big[h_{1T}^{(1/2) + g}\Big]^2 
   \,-\, \Big[h_{1T}^{(1/2) - g}\Big]^2 = 0 \label{SR3_sum} \,.
\en 
The first SR~(\ref{SR1_sum}) is fully consistent 
with the model-independent positivity bounds~(\ref{MR_bounds}) 
derived in Ref.~\cite{Mulders:2000sh}. Moreover, these bounds 
directly follow from this SR. 
 
At specific limits for $x$, these SRs get simplified and 
are consistent with model-independent predictions at small $x$, derived 
in Refs.~\cite{Boer:2015pni,Boer:2016xqr},  
\eq 
R_0 &=&  \Big[h_1^{(1) \perp g}\Big]^2 \,-\, \Big[f_1^g\Big]^2  = 0\,, 
\label{SR1_vlow_sum}\\
R_1 &=&  f_1^g   \, h_{1T}^{(1/2) - g}
   \,-\, h_1^{(1) \perp g} \, f_{1T}^{(1/2) \perp g} = 0\,, 
\label{SR2_vlow_sum}\\
R_2 &=&  \Big[f_{1T}^{(1/2) \perp g}\Big]^2  
   \,-\, \Big|h_{1T}^{(1/2) - g}\Big]^2 = 0 
\label{SR3_vlow_sum} 
\en 
and at large $x$, derived in LF QCD motivated by AdS/QCD~\cite{Lyubovitskij:2020xqj}, 
\eq 
R_0 &=&  \Big[g_{1L}^{(1) \perp g}\Big]^2 
   \,-\, \Big[f_1^g\Big]^2 
   = 0\,, \label{SR1_vhigh_sum}\\
R_1 &=&  f_1^g    \, \Big[h_{1T}^{(1/2) - g}
   \,+\, h_{1T}^{(1/2) + g} \Big] 
   = 0\,, \label{SR2_vhigh_sum}\\
R_2 &=&  
         \Big[h_{1T}^{(1/2) + g}\Big]^2 
   \,-\, \Big[h_{1T}^{(1/2) - g}\Big]^2 = 0 \label{SR3_vhigh_sum} \,.
\en 
From the SRs $R_1=0$ and $R_2=0$ at large $x$, we derive the constraint 
$h_{1T}^{(1/2) + g} = - h_{1T}^{(1/2) - g}$. 
 
In the case of T-even TMDs, our findings in LF QCD are summarized as~\cite{Lyubovitskij:2020xqj} 

\begin{itemize}

\item[(i)]
Large $x$ scaling: 
$f_1^g(x,\bfk^2)$ and $g_{1L}^g(x,\bfk^2)$ TMDs scale as $(1-x)^4$, 
while $g_{1T}^g(x,\bfk^2)$ and $h_1^{\perp g}(x,\bfk^2)$ TMDs have an
extra power $(1-x)$ falloff. 

\item[(ii)]
MR inequalities (positivity bounds): 
analytically fulfilled without referring to a specific 
choice of LFWFs.

\item[(iii)]
All four TMDs are expressed in terms of unpolarized $G(x)$ and 
polarized $\Delta G(x)$ PDFs. 

\item[(iv)]
A set of leading-twist gluon TMDs, based on the LF QCD formalism, was derived 
[see Eq.~(\ref{f1g_QCD})]. Our results are consistent 
with the model-independent MR inequalities involving gluon TMDs, 
with model-independent results for small $x$ behavior of gluon TMDs, 
and with constituent counting rules at large $x$. 
Taking into account the large $x$ behavior, we can suggest an improvement 
for the modeling of T-even TMDs. Our results for the T-even gluon TMDs 
are summarized in Eqs.~(\ref{f1g_QCD})-(\ref{rho}). 
 
\end{itemize}

In the case of T-odd gluon TMDs, we propose to express them in terms of T-even TMDs 
multiplied with four scalar functions depending on $x$ and $\bfk^2$ 
[see Eqs.~(\ref{h1Lx0})-(\ref{f1Tx0})]. 
Note that we derived the constraints on the small/large $x$ behavior of these 
four functions using model-independent constraints on the T-odd gluon TMDs. 
Our results for the T-odd gluon TMDs are summarized in Eqs.~(\ref{h1Lg})-(\ref{f1Tx0}).  
We hope that the findings in the present manuscript will be useful 
for more accurate extractions of gluon TMDs from data in running 
and planning experiments on hard-scattering high $p_T$ reactions ---
direct photon, light hadron, and heavy quarks production and 
taking into account the SRs between them and their large $x$ behavior. 
Note that in the derivation of large $x$ behavior of gluon 
PDFs/TMDs, in the present paper we rely on the results of Refs.~\cite{Brodsky:1989db,Brodsky:1994kg}. 
However, one should take into account the results of Ref.~\cite{Avakian:2007xa} 
motivated and verified by experimental studies 
at JLab~\cite{Zheng:2003un,Dharmawardane:2006zd,Parno:2014xzb} 
for a modification of the large $x$ behavior of the quark PDFs 
by large logarithmic corrections induced by the quark 
nonzero orbital angular momentum. One could expect 
a similar modification in the case of 
gluon PDFs, since gluons are also expected to carry the orbital angular momentum. 
We plan to study such effects of modifications of the gluon PDFs, 
which serve as input parameters in our parametrization for the gluon TMDs. 

\begin{acknowledgments}

This work was funded by BMBF ``Verbundprojekt 05P2018 - Ausbau von ALICE                                      
am LHC: Jets und partonische Struktur von Kernen''
(F\"orderkennzeichen: 05P18VTCA1), by ANID PIA/APOYO AFB180002 (Chile), 
and by FONDECYT (Chile) under Grants No. 1191103 and No. 1180232.  

\end{acknowledgments}

\end{document}